\documentclass[pdftex,superscriptaddress,preprint]{revtex4-2}

\usepackage{graphicx}
\usepackage{amsmath} 
\usepackage{amssymb}
\usepackage{physics}
\usepackage{amsthm}
\usepackage{mathtools}
\usepackage{bm}
\usepackage{gensymb}
\usepackage{wasysym}
\usepackage{mathtools}
\usepackage{braket}
\usepackage{upgreek}
\usepackage{cases}
\usepackage{rotating}
\usepackage{subfig}
\usepackage{caption}
\usepackage{tabularx}
\usepackage{float}
\usepackage[colorlinks=true,bookmarks=false,linkcolor=red,citecolor=red,urlcolor=red]{hyperref}

\renewcommand{\vec}[1]{{\mathbf{\bm{#1}}}}
\newcommand{\kk}{{\vec{k}}} 
\newcommand{\KK}{{\vec{K}}} 

\newcommand{\uv}[1]{\ensuremath{\hat{\mathbf{#1}}}} 

\renewcommand{\aa}{{\vec{A}(t)}}

\begin{document}

\title{Disentangling High Harmonic Generation from\\
Surface and Bulk States of a Topological Insulator}

\author{Sha Li}
\email{li.8187@osu.edu}
\affiliation{Department of Physics, The Ohio State University, Columbus, Ohio 43210, USA}

\author{Wenyi Zhou}
\affiliation{Department of Physics, The Ohio State University, Columbus, Ohio 43210, USA}

\author{Kazi A. Imroz}
\affiliation{Department of Physics, The Ohio State University, Columbus, Ohio 43210, USA}

\author{Yaguo Tang}
\altaffiliation{Present address: SLAC National Accelerator Laboratory, Menlo Park, California 94025, USA}
\affiliation{Department of Physics, The Ohio State University, Columbus, Ohio 43210, USA}

\author{Tiana A. Townsend}
\affiliation{Department of Physics, The Ohio State University, Columbus, Ohio 43210, USA}

\author{Vyacheslav Leshchenko}
\affiliation{Department of Physics, The Ohio State University, Columbus, Ohio 43210, USA}

\author{Larissa Boie}
\affiliation{Department of Physics, The Ohio State University, Columbus, Ohio 43210, USA}

\author{Pierre Agostini}
\affiliation{Department of Physics, The Ohio State University, Columbus, Ohio 43210, USA}

\author{Alexandra S. Landsman}
\affiliation{Department of Physics, The Ohio State University, Columbus, Ohio 43210, USA}

\author{Roland K. Kawakami}
\affiliation{Department of Physics, The Ohio State University, Columbus, Ohio 43210, USA}

\author{Lun Yue}
\email{lyue2@binghamton.edu}
\affiliation{Department of Physics, Applied Physics and Astronomy, Binghamton University, Binghamton, New York 13902, USA}

\author{Louis F. DiMauro}
\affiliation{Department of Physics, The Ohio State University, Columbus, Ohio 43210, USA}


\begin{abstract}
	
The discovery of topological phases has introduced a new dimension to materials science. Three-dimensional (3D) topological insulators (TIs) are a remarkable class of matter that is insulating in the bulk while hosting conductive topological surface states (TSSs) with unique charge and spin properties. High-order harmonic generation (HHG) has emerged as a powerful tool to probe condensed matter systems by providing insights into their electronic structure and dynamic behavior. Here, we investigate HHG in the prototype 3D-TI Bi$_2$Se$_3$. We demonstrate that the contributions of bulk and surface states to the harmonic emission can be controlled by tuning the thickness of thin film samples. An ultrathin (6 nm) film substantially enhances HHG from the surface states, while the bulk states dominate HHG in a thicker (50 nm) film. By applying a quasi-static terahertz perturbing field, we disentangle the bulk and surface responses and reveal the significant impact of the surface states' shift vector and Berry curvature on HHG. Our study provides effective methods for isolating the optical responses of TSSs from those of the bulk, which opens the door to resolving an ongoing debate regarding whether it is possible to reliably extract topological signatures in HHG.

\end{abstract}

\maketitle

\newpage
\section*{Introduction}
In the 1980s, foundational concepts related to topological phenomena, particularly in the context of the quantum Hall effect, were established \cite{IQHEKlitzing,FQHELaughlin,QHETheo}, leading to a new paradigm for classifying materials based on their topological order. Three-dimensional (3D) topological insulators (TIs) \cite{TIRevHasanKane,TIRevZhang} are a quantum phase of matter that is insulating in the bulk but conductive on the surface. The metallic surface arises from the presence of gapless Dirac cone topological surface states (TSSs), which are inherently tied to the non-trivial topology of the bulk band structure that experiences a band inversion under time-reversal symmetry (TRS) and strong spin-orbit coupling (SOC).The TSSs are robust against time-reversal-preserving (e.g., non-magnetic) perturbations and exhibit unique properties \cite{TIRevHasanKane,TIRevZhang}: they are spin-polarized, and the spin and momentum of the charge-carriers are locked at a right angle (spin-momentum locking). Additionally, their helical spin texture supports chiral spin current, the spin-momentum locking prevents back-scattering allowing for dissipationless charge and spin transport, and the spin vortex induces a $\pi$ Berry phase \cite{BerryPhase} in the wavefunction when looping around the Dirac point. These properties make 3D-TIs of substantial fundamental and practical interest.

Methods for exploring TIs include (spin and) angular resolved photoelectron spectroscopy (s-ARPES) \cite{TIExpVerBiSb,TIExpVerBiTe,TIExpVerBiSe}, measurement of the (quantum) Hall effect \cite{QSHEHgTe,QAHEMTI}, and circular dichroism examination \cite{CDBiSe}, which are employed to probe intrinsic properties of TIs or their perturbative interactions with light. Recently, high-order harmonic generation (HHG) has been observed in 3D-TIs \cite{LiuHHGBSTS,HuberHHGBiTe,HeideHHGBiSe}, offering an all-optical probe of them interacting with strong fields. HHG is a frequency up-conversion process that occurs when a strong laser field interacts with matter, typically atomic or molecular gases \cite{LHuillierHHG1988}. It is a hallmark of ultrafast physics and a key method for generating extreme-ultraviolet (XUV) and X-ray photons as well as attosecond pulses. The realization of HHG in solid-state systems \cite{GhimireHHGZnO} has launched extensive efforts to uncover ultrafast charge-carrier dynamics in strong-field-driven solids and to exploit high-harmonic spectroscopy as a probe of condensed-matter properties \cite{RiesHHGRev}.

The ability of HHG to probe topology has recently become a topic of animated debate. On one hand, a number of influential works have highlighted intriguing features in the HHG spectra of materials with non-trivial topology. For instance, the carrier-envelope-phase-dependent, non-integer “harmonic” generation from TSSs has been explained by the ballistic acceleration of Dirac fermions through the vicinity of the Dirac point \cite{HuberHHGBiTe,HuberTIBallAcc}. The “anomalous” (non-gas-like) dependence of the harmonic yield on the driving laser ellipticity has been attributed to the vortex spin texture and the hexagonal warping of the TSSs \cite{HeideHHGBiSe,Baykusheva2021PRA}. Calculations of modeled systems have predicted orders of magnitude enhancement in harmonic conversion efficiency \cite{HHGModelTP1} and sign flipping of harmonic light helicity \cite{HHGModelTP2} between the non-trivial and trivial topological phases. On the other hand, a subsequent work has suggested that the non-integer harmonics were due to the chirp of the driving pulse \cite{NonInterHHGChirp}. Anomalous HHG driven by elliptically or circularly polarized light has also been observed in many topologically trivial materials \cite{AnorHHGMgO,AnorHHGGaSe,AnorHHGGraphene} and hence alone may not reflect the system's topology. Furthermore, a numerical study cast doubt on the reliability of extracting topological phases via HHG from prior works \cite{RubioHHGTI}.

A serious challenge in probing TSSs via HHG is suppressing the responses of bulk states. Due to the very small bulk band gap of TIs, charge-carrier excitation and harmonic generation in the bulk can easily be triggered by a strong laser field. In addition,  typically, mid-infrared (MIR) and long-wavelength infrared (LWIR) pulses are used to drive HHG in TIs, producing harmonic light in the visible to infrared wavelength regime. The absorption lengths of the fundamental and harmonic light are respectively $>$100 nm and $\sim$20-30 nm \cite{TIabsorp}, while the TSSs only have a penetration depth of less than 2 nm \cite{TSSdepth}. Therefore, even in reflection geometry, the contribution of bulk states to the harmonic emission can be significant, particularly when only in-plane responses are considered.

Nevertheless, experiments can be carefully designed to selectively probe HHG from the surface states. For the Bi$_2$Se$_3$ family of TIs, the bulk belongs to the inversion symmetric $\mathrm{D}_{3\mathrm{d}}$ point group while the surface symmetry reduces to $\mathrm{C}_{3\mathrm{v}}$. As a result, even-order harmonics arising from electric-dipole transitions are permitted only from the surface, with their relative efficiency serving as a potential indicator of the contribution of surface states to HHG. Ref.~\cite{HuberHHGBiTe} suggests that by using a weak below-band gap driving field, HHG in the bulk of Bi$_2$Te$_3$ can be strongly suppressed. The spectrum consisting of commensurate odd- and even-order harmonic light has been attributed to the TSSs. However, for Bi$_2$Se$_3$ and BiSbTeSe$_2$ (BSTS), most previous studies have reported the yields of even-order harmonics that are 2-3 orders of magnitude weaker than those of the odd-orders \cite{HeideHHGBiSe,LiuHHGBSTS}, suggesting that harmonic emission from the surface states is weak.

While most previous studies either assumed that HHG originates solely from the surface or lacked solid evidence to support this claim\textemdash particularly for odd-order harmonics, which are allowed from both the bulk and surface states\textemdash this manuscript presents a novel approach for studying HHG in the prototype 3D-TI Bi$_2$Se$_3$ and addresses the following key question: How can we suppress/enhance harmonic emission from the bulk/surface states and effectively distinguish between these contributions? We demonstrate: 1. The contributions of bulk and surface states to the MIR-driven HHG can be controlled by tuning the thickness of thin film samples.  An ultrathin (6 nm) film substantially enhances HHG from the surface states, while the bulk states dominate HHG in a thicker (50 nm) film. 2. Introducing a quasi-static terahertz (THz) perturbing field, i.e., a MIR-THz two-color field scheme, provides an unambiguous method to differentiate between the bulk and surface HHG responses, applicable to both the even- and odd-order harmonic light. Our study provides effective methods for isolating TSSs from the bulk, paving the way for further exploration of the role of non-trivial topology in HHG.

\begin{figure}[H]
	
	\centering
	\resizebox{\textwidth}{!}{\includegraphics{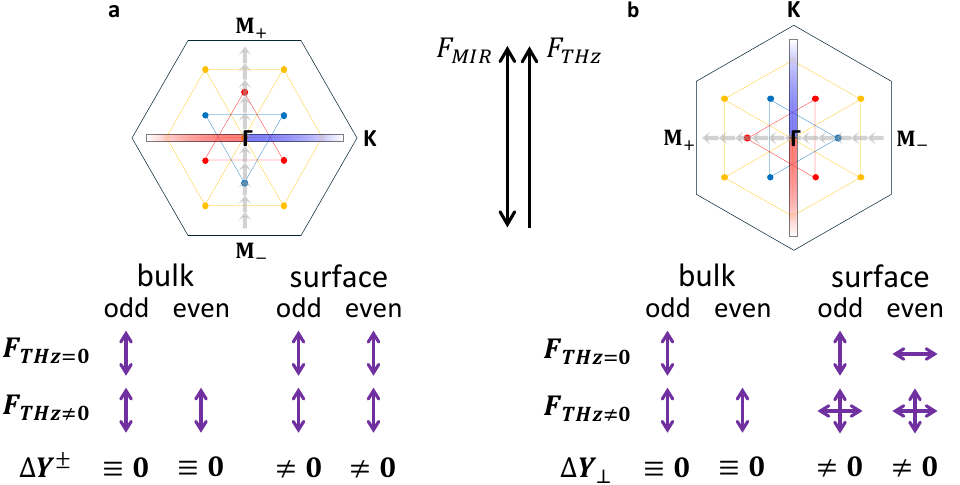}}
	\caption{\textbf{The experimental design.} Polarization-resolved harmonic selection rules when the driving fields are polarized along the real-space directions corresponding to the {\bf (a)} $\Gamma$M  or {\bf (b)} $\Gamma$K directions in reciprocal space. Black hexagons indicate the Bi$_2$Se$_3$ surface Brillouin zone, colored triangles with dots show the real-space lattice structure projected onto the (111) plane. Symmetry breaking of the TSSs is ``visualized" by the gray arrows and color-scale bar, which represent local directions of the in-plane shift vector $\mathbf R^{\kk,y}$ along $\Gamma$M and amplitude of the out-of-plane ($z$) Berry curvature $\mathbf \Omega^{\kk}_z$ along $\Gamma$K, respectively. Reciprocal space two-dimensional distributions of $\mathbf R$ and $\mathbf \Omega$ can be found in Fig.\@~\ref{fig2}. Black arrows indicate the MIR and (instantaneous) THz field directions. Purple arrows show the HHG polarization directions. {\bf (a)} When the driving fields are polarized along the $\Gamma$M direction, the differential HHG yield between the cases where the instantaneous THz field and the TSSs' shift vector are parallel ($+$, same direction, as shown in Fig.\@~a) or anti-parallel ($-$, opposite directions, 180$\degree$ crystal orientation of Fig.\@~a), defined as $\Delta Y^{\pm} \equiv Y(F_{\mathrm{THz} \,\parallel\, \Gamma \mathrm{M}_+})-Y(F_{\mathrm{THz} \,\parallel\, \Gamma \mathrm{M}_-})$, serves to distinguish harmonic emission from the bulk ($\mathbf{R} = 0$, $\Delta Y = 0$) and surface ($\mathbf{R} \neq 0$, $\Delta Y \neq 0$) states. {\bf (b)} When the driving fields are polarized along the $\Gamma$K direction, and for the harmonic emission perpendicular to the driving fields polarization, the HHG yield difference with and without the THz field, defined as $\Delta Y_{\perp} = Y_{\perp}(F_{\mathrm{THz} \neq 0})-Y_{\perp}(F_{\mathrm{THz}=0})$,  probes an exclusive surface states HHG response (Bulk: $\mathbf{\Omega} = 0$, $\Delta Y_{\perp} = 0$; Surface: $\mathbf{\Omega} \neq 0$, $\Delta Y_{\perp} \neq 0$).}
	\label{fig1}
	
\end{figure}

Symmetry is important throughout our discussion. Henceforth we will use $\mathcal P$- and $\mathcal T$- to denote parity (inversion) and time-reversal symmetry, respectively. For convenience, the real-space crystal orientation is described using the corresponding reciprocal-space symmetry directions $\Gamma$M or $\Gamma$K (see Fig.\@~\ref{fig1}).

\section*{Results and Discussion}
The experiments are conducted with normal incident light in transmission geometry. A strong 60 fs, 3.6 $\mu$m MIR pulse drives HHG in the (111) plane of Bi$_2$Se$_3$ thin-film crystal grown on (0001) Al2O3 substrate. A weak $\sim$2 ps period single-cycle THz pulse is applied to perturb or control the HHG process. The THz field is considered quasi-static since its period is much longer than the MIR pulse duration. Both the MIR and THz fields are linearly polarized along the vertical direction in the laboratory frame, and the crystal can be oriented with respect to the laser polarization. Harmonic light is dispersed by a monochromator and detected by an ICCD camera. The harmonic polarization is analyzed using a Rochon prism.  Detailed experimental setup can be found in the ``Methods" section. 

\subsection*{Experimental concept}
According to Floquet group theory and electric dipole selection rules, the forbidden even-order HHG in a $\mathcal{P}$-invariant medium driven by a monochromatic laser field arises from the $H(\textbf{r},t)=H(-\textbf{r},t+T_0/2)$ dynamical symmetry of the Hamiltonian \cite{HHGDySym1,HHGDySym2}, here $T_0$ is the laser period. An intrinsic $\mathcal P$-violation of the medium or an asymmetric driving field $\vec{F}(t) \neq - \vec{F}(t+T_0/2)$ both can break this symmetry and promote even-order harmonic emission. Our experiments exploit this principle, we introduce an asymmetry in the driving field by applying a quasi-static THz perturbation, and demonstrate that the coupling between intrinsic and THz-field-induced symmetry breaking enables the differentiation of surface ($\mathcal P$-violated) and bulk ($\mathcal P$-invariant) HHG responses in Bi$_2$Se$_3$. 

Intrinsic $\mathcal P$-violation that gives rise to even-order optical responses is often characterized by vector quantities. For instance, non-zero Berry curvature ($\mathbf \Omega$, pseudovector) and shift vector ($\mathbf R$, polar vector) due to $\mathcal P$-violation are crucial for the generation of even-order harmonics. Figure\@~\ref{fig2} shows $\mathbf \Omega$ and $\mathbf R$ for the TSSs of Bi$_2$Se$_3$. Phenomenologically, they can be understood as vectors that define the directions of symmetry breaking in the crystal frame, similar to the spin of spin-polarized molecules or the polar axis of polar molecules. In condensed matter physics, $\mathbf \Omega$ plays an essential role in phenomena such as the anomalous Hall effect \cite{AHEHall,AHE}, while $\mathbf R$ contributes to the shift current photovoltaic effect \cite{ShiftVec1,ShiftVec2,ShiftVec3}. The roles of $\mathbf \Omega$ and $\mathbf R$ in strong-field HHG have been actively studied recently \cite{LYAnoHHG,ShiftVectorHHG,GISBEs}, and a brief discussion is provided in Supplementary Information (SI) Section 3. Quantitative analysis of HHG with non-zero $\mathbf \Omega$ and $\mathbf R$ not only offers a means to extract intrinsic material properties but also deepens our understanding of the HHG dynamics in solids \cite{BerryCurvatureHHG1,BerryCurvatureHHG2,BerryPhaseTunnelingHHG,BerryPhaseEffHHG}.

In this study, we demonstrate that the bulk ($\mathbf \Omega =0$, $\mathbf R = 0$) and surface ($\mathbf \Omega \neq 0$, $\mathbf R \neq 0$) states of Bi$_2$Se$_3$ produce distinct harmonic spectra, and respond differently to the THz perturbation. Moreover, the interplay between intrinsic and THz-field-induced symmetry breaking in TSSs enables us to probe the roles of the shift vector and Berry curvature in HHG. The expected experimental outcomes are summarized in Fig.\@~\ref{fig1}. To support the key features of our experimental observations, we perform time-dependent strong-field simulations using a tight-binding model for the TSSs combined with a gauge-invariant formulation of the semiconductor Bloch equations (see ``Methods" for simulation details and SI Section 4 for simulation results).

\subsection*{Control HHG from the bulk and surface states by thin film thickness}
The experiments that compare the bulk- and surface-dominated HHG responses are realized with high quality thin film samples of varying thicknesses. Given the extremely short penetration depth ($\sim$2 nm) of the TSSs, we fabricate an ultrathin film with a thickness of 6 nm ($\approx$1 nm per layer) such that it primarily consists of two (top and bottom) surfaces while the bulk volume is minimized. This thickness is chosen because studies have shown that when the film is thinner than 5-6 nm, the interaction between top and bottom TSSs results in re-opening of the band gap and the material may lose its non-trivial topology \cite{TSSTopBottomInterac1}. For comparison, we also prepare a thick film of 50 nm thickness to represent a sample with maximized effective bulk volume. Films thicker than 50 nm do not further enhance the bulk HHG due to the short absorption length of the harmonic light. 

Figure\@~\ref{fig3} shows the harmonic spectra for the 6 and 50 nm samples. Due to the longer effective interaction length for HHG, the total harmonic yield for the 50 nm sample is a few times stronger than that for the 6 nm sample. However, the even-order harmonics, which can only be generated from the $\mathcal P$-violated surface, are nearly two orders of magnitude stronger in the 6 nm sample compared to the 50 nm sample, suggesting that the surface states' HHG is significantly enhanced in the ultrathin film. When the MIR driving field is polarized along the $\Gamma$K direction, the yields of adjacent odd and even harmonic orders are commensurate, similar to the LWIR-driven HHG in Bi$_2$Te$_3$ in reflection geometry \cite{HuberHHGBiTe}. Such efficient relative even-order harmonic conversion strongly indicates that in the ultrathin film, HHG is dominated by the surface states.

To investigate how HHG from the bulk and surface states evolves with the film thickness, we measure the harmonic yields for samples of varying thickness. The results are shown in Fig.\@~\ref{fig4}. The total harmonic yield increases with the film thickness and peaks at about 20 nm. This saturation behavior is attributed to the short absorption lengths of both the fundamental and harmonic light, as well as thin film interference and self-phase modulation of the fundamental light which may affect the laser peak intensity. In contrast, the total yield of surface-exclusive even-order harmonics increases as the film thickness decreases, with a steep rise below 10 nm. This sharp transition and significant increase of surface HHG in films thinner than 10 nm cannot be explained by light absorption alone. It indicates that the reduced bulk volume plays a central role in enhancing surface HHG and points to a mechanism in which the bulk and surface of a TI are not fully isolated. We attribute this mechanism to bulk-surface electron scattering \cite{BSScatterSTM,BSScatterTrARPES}, and provide a detailed discussion in SI Section 1. In brief, the photoexcited bulk electrons scatter into TSSs, incoherently generating excess hot electrons and increasing the electron-electron scattering within the surface during HHG, thereby suppressing the efficiency of surface HHG \cite{DopingHHGZnO,DopingHHG}. Because bulk states are delocalized throughout the crystal while TSSs are confined within only $\sim$2 nm near the surface, only bulk electrons sufficiently close to the surface scatter efficiently into TSSs. The observed transition near 10 nm therefore suggests an upper bound of $\sim$5 nm for the bulk-surface coupling length (Supplementary Fig.\@~S1).

\subsection*{Disentangling HHG from the bulk and surface states with a terahertz field}
In this section, we confirm the dominant contributions of bulk and surface states to the HHG from the 50 nm and 6 nm samples, respectively, and verify the experimental concept introduced earlier: by applying a quasi-static terahertz perturbing field, the bulk and surface HHG responses are disentangled, allowing us to examine the influence of TSSs' shift vector ($\textbf R$) and Berry curvature ($\mathbf \Omega$) on HHG. 

\begin{figure}[H]
	
	\centering
	\resizebox{\textwidth}{!}{\includegraphics{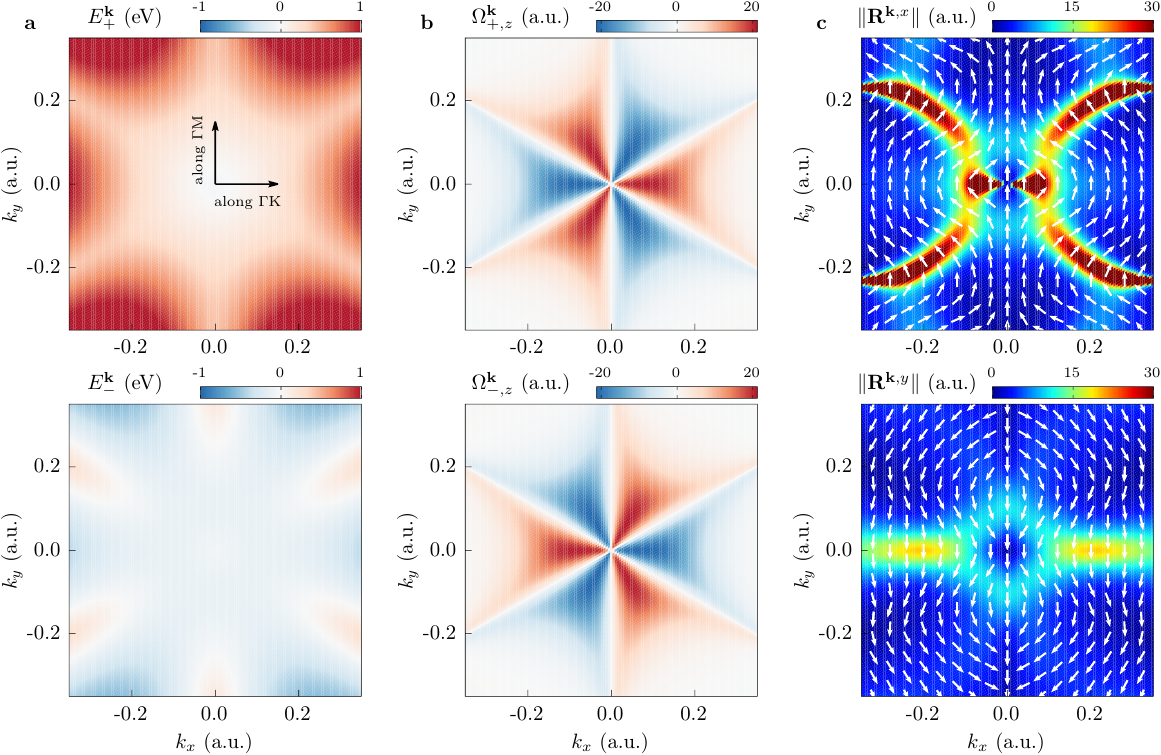}}
	\caption{\textbf{Reciprocal space inversion symmetry breaking in the (111) plane of Bi$_2$Se$_3$ TSSs.} {\bf (a)} Band structure $E^{\kk}$ and {\bf (b)} amplitude of the out-of-plane Berry curvature $\mathbf \Omega^{\kk}_z$, for the Dirac cone upper (``$+$", top panel) and lower (``$-$", bottom panel) states. {\bf (c)} Shift vector $\mathbf{R}^{\kk, x} \equiv (R_{xx}^\kk, R_{yx}^\kk)$ (top panel) and $\mathbf{R}^{\kk, y} \equiv (R_{xy}^\kk, R_{yy}^\kk)$ (bottom panel); the color scale and white arrows indicate the magnitude and local direction of the shift vector, respectively. $\mathcal T$-invariance requires a symmetric band structure $E^\kk=E^{-\kk}$. The surface states $\mathcal P$-violation is restored by the non-zero $\mathbf \Omega$ and $\mathbf R$, which are odd and even functions of $\kk$, respectively: $\mathbf{\Omega}^\kk=-\mathbf{\Omega}^{-\kk}$ and $\mathbf{R}^\kk = \mathbf{R}^{-\kk}$. See Equations 4 and 6 in ``Methods" for the mathematical definitions, and Section 3 in SI for detailed physical meanings, of $\mathbf{\Omega}$ and $\mathbf{R}$. }
	\label{fig2}
	
\end{figure}

\begin{figure}[H]
	
	\centering
	\resizebox{\textwidth}{!}{\includegraphics{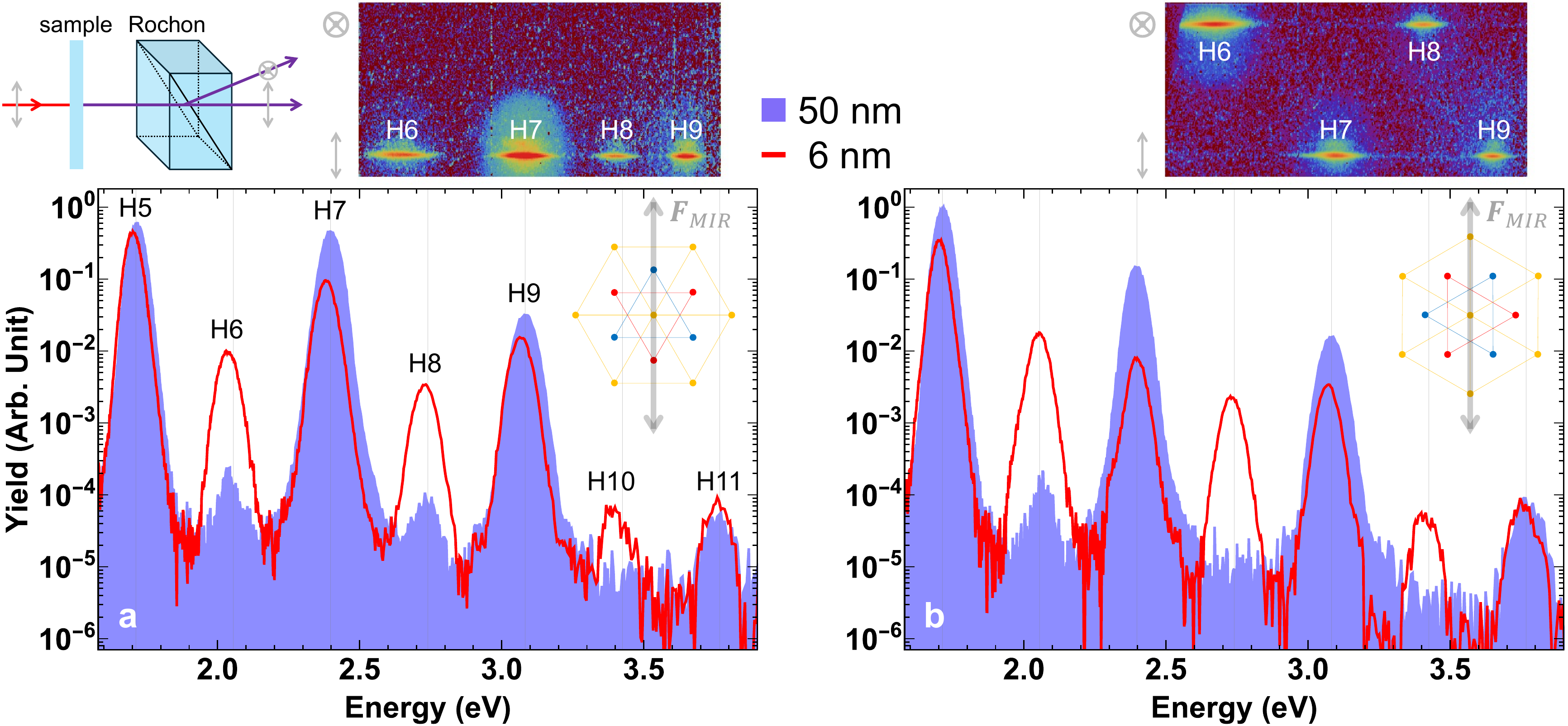}}
	\caption{\textbf{MIR-driven high-order ($5^\mathrm{th}$ to $11^\mathrm{th}$) harmonic spectra from the Bi$_2$Se$_3$ thin film crystals.} The 50 nm (shaded blue) and 6 nm (solid red) samples are oriented with {\bf (a)} $\Gamma$M and  {\bf (b)} $\Gamma$K along the MIR polarization direction, respectively. All four spectra are scaled by the same factor. {\bf Top:} Schematics of the polarization-resolved HHG measurements using a Rochon prism: harmonic components polarized parallel and perpendicular to the fundamental field are spatially separated along the vertical direction and simultaneously detected by the spectrometer camera. Gray arrows indicate the polarization direction of the light.}
	\label{fig3}
	
\end{figure}

\begin{figure}[H]
	
	\centering
	\resizebox{\textwidth}{!}{\includegraphics{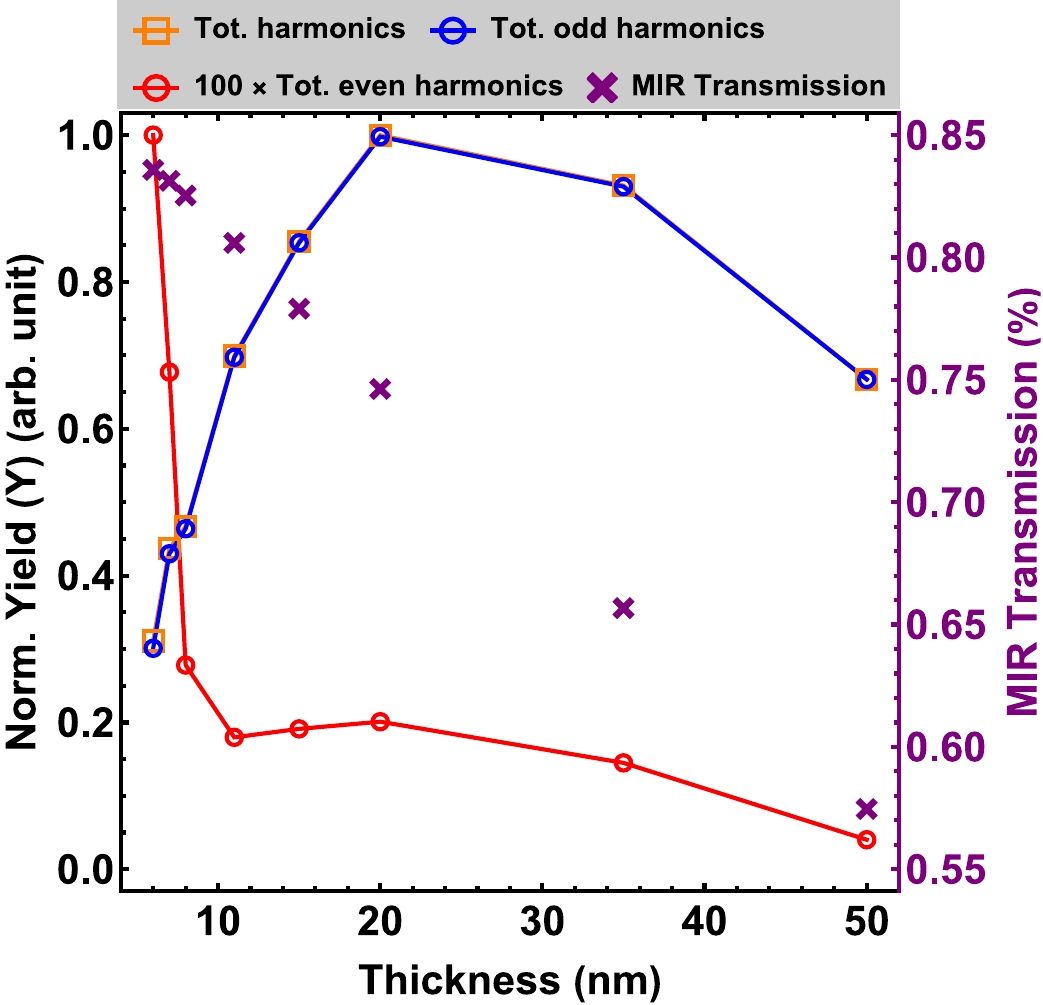}}
	\caption{\textbf{Evolution of the harmonic yield (Y) and MIR transmission as a function of the thin film thickness.} Total harmonic yield $\sum\limits_{q=5}^{11} \textrm{Y}_{q}$ (open orange squares), total odd-order harmonic yield (open blue circles), 100 $\times$ total even-order harmonic yield (open red circles). All three curves are normalized by the same factor (the total harmonic yield maximum). The MIR driving laser transmission is shown as purple crosses. Despite of large ($>$5) index of refraction, the MIR transmission is high due to thin film interference.}
	\label{fig4}
	
\end{figure}

\begin{figure}[H]
	
	\centering
	\resizebox{\textwidth}{!}{\includegraphics{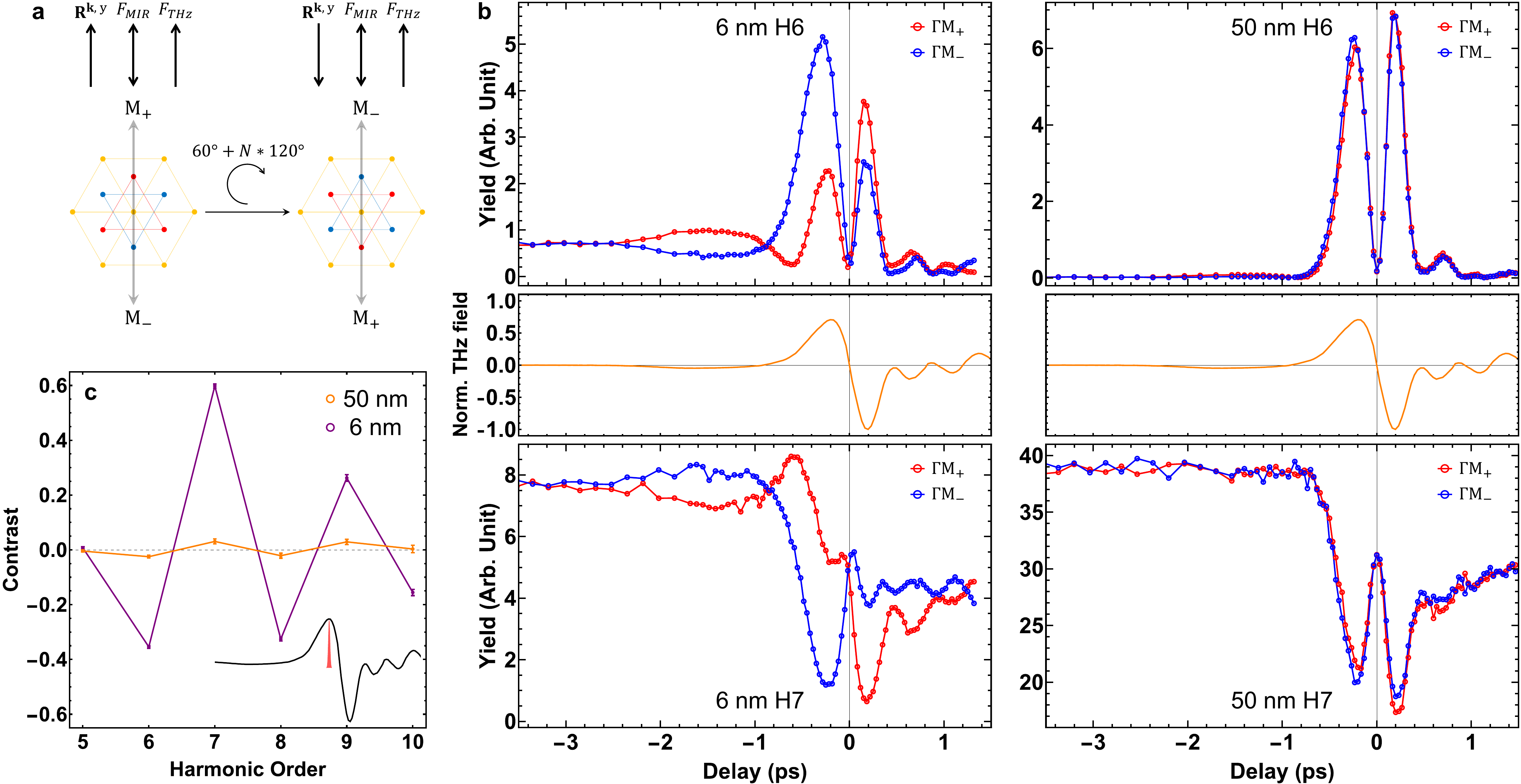}}
	\caption{\textbf{Probing the influence of shift vector on HHG by the THz field.} {\bf (a)} Experimental configuration: The MIR and THz fields are polarized along the $\Gamma$M direction. In the crystal frame, M$_+$ and M$_-$ are defined by the polarity of the TSSs' shift vector $\mathbf R^{\kk,y}$ along $\Gamma$M, which points from M$_-$ to M$_+$. In the laboratory frame, the polarity of the first main half-cycle of the THz field (see inset {\bf c}) defines the ``up" direction.  {\bf (b)} For the 6 nm (left panels) and 50 nm (right panels) samples, the $6^\mathrm{th}$- (upper panels) and $7^\mathrm{th}$- (lower panels) order harmonic yields are measured as a function of the MIR-THz time delay, for the two crystal orientations shown in {\bf a} when $\Gamma$M$_+$ (red) and $\Gamma$M$_-$ (blue) point upwards. All yields are scaled by the same factor. Time zero is defined at the field zero-crossing of the THz main cycle, and large negative time delays correspond to MIR leading. Middle panels show a typical THz waveform at focus in air, measured separately. {\bf (c)} At a fixed MIR-THz time delay of about $-250$ fs when the MIR pulse overlaps with the crest of the first main half-cycle of the THz field (see inset), the contrast between $\Gamma$M$_+$ and $\Gamma$M$_-$ crystal orientations, defined as the harmonic-yield difference divided by their sum, $(Y_{+} - Y_{-})/(Y_{+} + Y_{-})$, is shown as a function of the harmonic order. Error bars are standard deviations of 10 measurements.}
	\label{fig5}
	
\end{figure}

As shown in  Fig.\@~\ref{fig2}, along the $\Gamma$M direction, $\mathbf \Omega$ vanishes and $\mathbf R$ exhibits only a longitudinal component, the polarity of $\mathbf R$ characterizes the direction of symmetry breaking in the crystal frame. We define M$_+$ and M$_-$ such that $\mathbf R^{\kk,y}$ points from M$_-$ to M$_+$ along $\Gamma$M. When the driving field is polarized along $\Gamma$M, dynamical symmetry restricts all the harmonic light to be polarized parallel to the driver (see Fig.\@~\ref{fig3}{\color{red} a-top} for measurement, and Supplementary Fig.\@~S5 for simulation). Due to $\mathcal T$-invariance, the band structure is symmetric regardless of bulk or surface states, therefore $\Gamma$M$_+$ and $\Gamma$M$_-$ are indistinguishable from the harmonic yield with a monochromatic (bipolar) driver. However, for the TSSs, the polarity of $\mathbf R$ can be referenced by applying a polar THz perturbing field, which serves as a direct knob to tune the effect of $\mathbf R$ on HHG. In the experiments, we orient the crystal to compare HHG in the two cases where $\mathbf R^{\kk,y}$ points towards ($\Gamma$M$_+$) or opposite ($\Gamma$M$_-$) to the instantaneous THz field direction, as shown in Fig.\@~\ref{fig5}{\color{red} a}. To the lowest order nonlinearity, this investigates the coupling between intrinsic and electric field-induced second harmonic generation (i-SHG and EFISHG \cite{EFISHG}). The difference between the two cases can be understood through perturbative nonlinear optics: $P_{2\omega}^{\pm} = (\chi^{(2)} \mp \chi^{(3)} F_\textrm{THz}) F_\omega^2$, here $P$, $F$, and $\chi$ are absolute values of the polarization, electric field, and susceptibility, respectively. Only the $\mathcal P$-violated ($\chi^{(2)} \neq 0$) surface states produce a difference in the harmonic yield ($\propto P^2$) between the two orientation configurations.

Figure\@~\ref{fig5}{\color{red} b} shows the $6^\mathrm{th}$- and $7^\mathrm{th}$-order harmonic yields as a function of the MIR-THz time delay. For the 6 nm sample, both harmonics exhibit pronounced differences between the $\Gamma$M$_+$ and $\Gamma$M$_-$ crystal orientations, while the differences are minimal for the 50 nm sample, indicating that these two harmonics are dominated by the surface and bulk states for the 6 and 50 nm samples, respectively. At a fixed MIR-THz time delay, and from the $5^\mathrm{th}$- to $10^\mathrm{th}$-order harmonics, Fig.\@~\ref{fig5}{\color{red} c} plots the contrast between the two crystal orientations, defined as the harmonic-yield difference divided by their sum, $(Y_{+} - Y_{-})/(Y_{+} + Y_{-})$. The contrast is a parameter reflecting the strength of surface states HHG as pure bulk responses yield a zero contrast. For the 50 nm sample, the contrast values remain very low across all the harmonic orders, suggesting a bulk-dominated HHG. For the 6 nm sample, large contrasts, a signature of surface-dominated HHG, are observed except for the $5^\mathrm{th}$-order harmonic. 

Our measurements demonstrate that the influence of $\mathbf R$ on HHG can be probed by introducing a weak THz perturbation. Importantly, a recent theoretical study of HHG in ZnO driven by a single-color field concluded that $\mathbf R$ has little effect on odd-order harmonics \cite{GISBEs}. In contrast, our results highlight the crucial role of the THz perturbation in revealing the otherwise hidden impact of $\mathbf R$ on odd-order harmonic emission. Our simulations qualitatively reproduce the strong contrasts observed for the low-order harmonics (Supplementary Fig.\@~S7), while for harmonics above the 9$^\mathrm{th}$-order the simulated contrasts become much smaller, indicating our current model does not fully capture the higher-order responses.

The extremely small contrast of the 5$^\mathrm{th}$-order harmonic indicates that, under our experimental conditions, it is bulk-dominated even in an ultrathin (6 nm) film with minimum bulk volume. Additional HHG measurements by varying the MIR-THz time delay and by rotating the crystal orientation, presented in Supplementary Fig.\@~S3, further support this conclusion. This counterintuitive behavior is likely due to strong bulk charge-carrier excitation induced by the relatively high fundamental photon energy ($\sim$0.34 eV) above the bulk band gap ($\sim$0.3 eV), which produces strong low-order harmonic emission from the bulk.

\begin{figure}[H]
	
	\centering
	\resizebox{\textwidth}{!}{\includegraphics{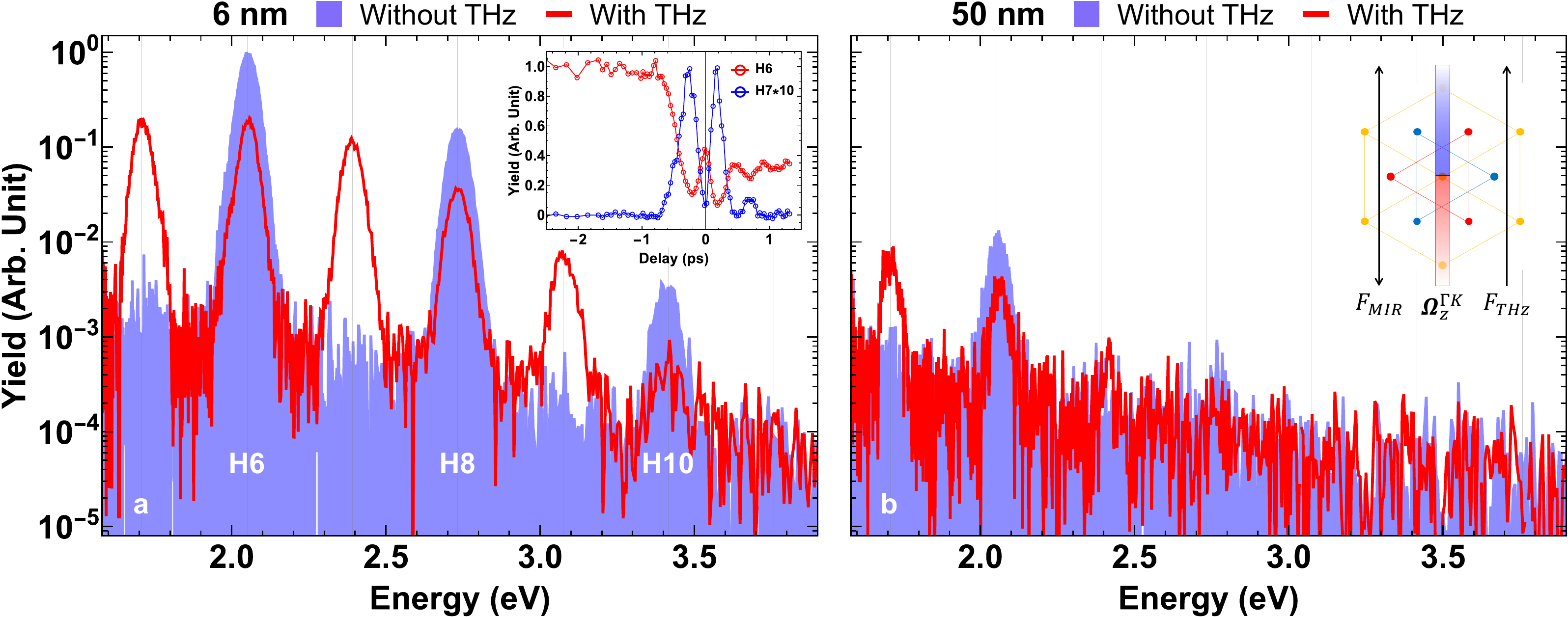}}
	\caption{\textbf{Probing the influence of Berry curvature on HHG by the THz field.} The MIR and THz fields are polarized along the $\Gamma$K direction, and only the harmonic light polarized perpendicular to the driving fields is measured.  For the {\bf (a)} 6 nm and {\bf (b)} 50 nm samples, perpendicular harmonic spectra without the THz field (shaded blue) and with the THz field (solid red) at a fixed MIR-THz time delay of about $-250$ fs (see Fig.\@~\ref{fig5}{\color{red} c} inset), are shown.  All four spectra are scaled by the same factor. {\bf Inset a}: For the 6 nm sample, the perpendicular $6^\mathrm{th}$- (red) and ($10\times$) $7^\mathrm{th}$- (blue) order harmonic yield as a function of the MIR-THz field time delay. {\bf Inset b}: Crystal orientation relative to the laser fields. The color-scale bar demonstrates amplitude of the out-of-plane Berry curvature along $\Gamma$K. } 
	\label{fig6}
	
\end{figure}

In two-dimensional and topological materials, the Berry-curvature-induced perpendicular HHG ($\abs{\bf{j}_{\perp,\Omega}} \propto \mathbf{F} \times \mathbf{\Omega}$) can be significant \cite{LYAnoHHG,HuberHHGBiTe}. Our approach provides a promising avenue to explore the Berry curvature effect on HHG by considering the case when the MIR driving field is polarized along the $\Gamma$K direction, where the longitudinal component of $\mathbf R$ vanishes and symmetry breaking is characterized by the out-of-plane $\mathbf{\Omega}$ (Fig.\@~\ref{fig2}). In this case, the Hamiltonian possesses a $\hat{Z}_\mathrm{M}=\hat{\tau}_2 \cdot \hat{\sigma}_v^{\mathrm{M}}$ reflection dynamical symmetry, restricting the odd- and even- (if allowed) order harmonic light to be polarized parallel and perpendicular to the fundamental driving field, respectively (see Fig.\@~\ref{fig3}{\color{red} b-top} for measurement, and Supplementary Fig.\@~S6 for simulation) \cite{HHGDySym1}. Here $\hat{\tau}_2$ and $\hat{\sigma}_v^{\mathrm{M}}$ are the time translation ($t \rightarrow t+T_0/2$) and mirror reflection (off the vertical mirror plane parallel to $\Gamma$M) operators. When a parallel THz field is applied, the $\hat{Z}_\mathrm{M}$ symmetry is lifted, affecting the HHG in both dimensions. However, the harmonic modulation in the perpendicular direction reflects an exclusive surface states response. We measure this effect in the 6 and 50 nm samples, as shown in Fig.\@~\ref{fig6}. In the perpendicular direction, only even-order harmonics are produced with the MIR driving field alone, applying the THz field not only suppresses the yields of even-order harmonics but also enables the generation of odd-order harmonics. Remarkably, for the surface-dominated 6 nm sample, the perpendicular harmonic yields exhibit pronounced modulations across the full spectrum, which are accurately captured by the simulated HHG spectra (Supplementary Fig.\@~S8), demonstrating the effectiveness of this method for probing HHG from isolated surface states.

In conclusion, we have demonstrated that the MIR-driven HHG from bulk and surface states of 3D-TI Bi$_2$Se$_3$ can be controlled by varying the thickness of thin film samples. The substantial enhancement of surface HHG in ultrathin films ($<$10 nm) is attributed to reduced bulk-surface electron scattering. To probe TIs via high harmonic spectroscopy, it is crucial to disentangle the bulk and surface responses. We achieve this by employing a MIR-THz two-color field scheme. The observed effects of the TSSs' shift vector and Berry curvature on HHG, probed by the THz field, are significant, making it intriguing to explore the potential role of non-trivial topology. Future experiments using ultrathin films with topologically trivial and non-trivial surface states (controllable through doping \cite{TIdoping}), in comparison with theoretical calculations, could provide deeper insights. In addition, the concepts introduced in this work\textemdash controlling the ratio of surface area to bulk volume by varying the thin film thickness, and exploiting the coupling between intrinsic and field-induced symmetry breaking to distinguish surface and bulk optical responses\textemdash are broadly applicable and will provide valuable guidance for future experimental designs of probing TSSs beyond HHG studies. Finally, we note that whether the non-trivial topological features of the bulk states, such as the band inversion, produce distinctive HHG signatures has largely been overlooked and merits future investigations.

\section*{Methods}
\subsection*{Experimental setup}
In the experiments, $\sim$60 fs (intensity FWHM), 3.6 $\mu$m (83.3 THz, 0.34 eV) MIR pulses are generated from an optical parametric amplifier based on KTiOAsO4 (KTA) crystals. Single-cycle $\sim$2 ps period (600 $\mu$m, 0.5 THz, 2.1 meV) THz pulses are generated via tilted-pulse-front-pumping (TPFP) optical rectification in a LiNbO$_3$ crystal. Both the MIR and the THz generations are pumped by laser pulses delivered from the same Ti:sapphire system (80 fs, 800 nm, 5.5 mJ, 1 kHz). The THz pulse characteristics (central frequency, asymmetry between positive and negative half-cycles, small satellite cycles before/after the main cycle) may have small day-to-day variations, however, these do not affect our study in this manuscript. The MIR and THz fields are linearly polarized in the laboratory frame vertical direction. The MIR intensity and polarization are controlled via a half-waveplate followed by a nano-particle polarizer (Thorlabs, extinction ratio (ER) $>10^{6}$), and the THz intensity and polarization are controlled by a pair of wire grid polarizers (Specac, ER $>10^{9}$). Harmonic light is collected by a visible-vacuum ultraviolet monochromator (VIS-VUV, 120-900 nm) and detected by a cooled ICCD camera (Princeton Instruments, 16 bits, noise level $<$5). The intensity standard deviation of the harmonic light is about 1\%. The harmonic polarization is analyzed by an ultrabroad-bandwidth (130-7000 nm) Rochon prism (Edmund optics, ER $>10^{5}$). The highly linearly polarized MIR and THz fields, high ER Rochon prism, and high dynamic range detector allow for high precision polarization measurements.

The experiments are conducted with normal incident light in transmission geometry in ambient air. The samples are mounted frontwards (the lasers hit the thin film first and then the substrate). The focal size ($1/e^2$ intensity diameter) of the MIR and THz beams are $\sim$220 $\mu$m and $\sim$1.5 mm, respectively. In the experiments, the maximum applied MIR and THz peak intensities (in air) are $\sim$80 GW cm$^{-2}$ and $\sim$0.48 GW cm$^{-2}$, respectively. We measure from the 5$^\mathrm{th}$- to 11$^\mathrm{th}$-order (327-720 nm) harmonic light. When pumping slightly above the crystal damage threshold, we are able to detect up to the 13$^\mathrm{th}$-order (277 nm) harmonic in Bi$_2$Se$_3$.

In this study, we only focus on the THz-field-induced symmetry breaking. However, we also notice that the extremely low-frequency THz pulse heats the crystal through incoherent charge-carrier acceleration. This heating lasts for several hundred picoseconds (verified by MIR-THz time delay scan), and globally reduces the HHG efficiency.

\subsection*{Sample preparation}
The highly-crystalline (111) Bi$_2$Se$_3$ thin films are deposited in a Veeco 930 molecular beam epitaxy (MBE) system with a base pressure of $2 \times 10^{-10}$ Torr. The 0.5 mm thick (0001) Al$_2$O$_3$ substrates (MTI Corporation), chosen for their favorable lattice and symmetry match, are sonicated in acetone, methanol, and isopropyl alcohol for five min each, and annealed in air at 1000 $\degree$C for three hours before loading into the MBE chamber. The substrates are then brought up to 800 $\degree$C in the ultra-high vacuum chamber to de-gas for an hour prior to growth. For synthesis, the films are grown by co-depositing Bi (99.95$\%$, Alfa Aesar) and Se (99.999$\%$, Puratronic) with an atomic flux ratio of $\sim$1:30. Substrate is first heated up to 260 $\degree$C to grow Bi$_2$Se$_3$ for 2 min, then annealed at 550 $\degree$C to evaporate it off. Substrate is then cooled down to 260 $\degree$C to deposit Bi$_2$Se$_3$ buffer layer for 2 min again before heated up to 360 $\degree$C to deposit until the desired thickness is reached. The pre-growth procedure of buffer layers is aimed to reduce twinning domains and surface roughness \cite{SampleGrowth}. All Bi$_2$Se$_3$ samples are capped with a 5 nm CaF$_2$ protection layer to prevent oxidization and degradation.

\subsection*{TSSs model and HHG simulation methods}

\noindent{\bf 1. Tight-binding model}

We consider a tight-binding model for the TSSs in Bi$_2$Se$_3$ \cite{Zhang2009NPhys,Shan2010NJP,Liu2010PRB,Baykusheva2021PRA} previously derived in Ref.~\cite{Baykusheva2021PRA}, which captures the correct symmetries of the TSSs and their properties such as spin-momentum locking and hexagonal warping \cite{Fu2009PRL} at large crystal momenta. 

The Hamiltonian of the TSSs can be parametrized to the form
\begin{equation}
	\label{eq:tb_1}
	H^\kk = \sum_{i=0}^3 B_i^\kk \sigma_i,
\end{equation}
with $\kk=(k_x,k_y)$ the in-plane crystal momenta, $\sigma_0\equiv I$ the identity matrix, $\{\sigma_i, i=1,2,3\}$ the Pauli matrices, and $\{B_i^\kk, i=1,2,3\}$ are functions defined in the tight-binding model \cite{Baykusheva2021PRA}. As usual, it is convenient to consider the system as an analog to a spin-$\tfrac{1}{2}$ particle in a pseudo-magnetic field, and define polar and azimuthal angles as \cite{Chacon2020PRB}
$\cos\theta_\kk = B_3^\kk/{\sqrt{\sum_{i=1}^3B_i^\kk}}$ 
and
$\phi_\kk = \arg(B_1^\kk + i B_2^\kk)$,
where $B^\kk\equiv \sqrt{\sum_{i=1}^3B_i^\kk}$.
The eigenstates can then be written in a particular gauge as
\begin{equation}
\label{eq:tb_3}
\ket{u_+^\kk} =
\begin{pmatrix}
	\cos\tfrac{\theta_\kk}{2} \\
	e^{i\phi_\kk} \sin\tfrac{\theta_\kk}{2}
\end{pmatrix}
,
\qquad
\ket{u_-^\kk} =
\begin{pmatrix}
	e^{-i\phi_\kk} \sin\tfrac{\theta_\kk}{2} \\
	-\cos\tfrac{\theta_\kk}{2}
\end{pmatrix}
.
\end{equation}
with the eigenenergies
\begin{equation}
\label{eq:ham_2}
E_{\pm}^\kk = B_0^\kk \pm B^\kk.
\end{equation}

Here ``$+$" and ``$-$" correspond to the Dirac cone upper and lower states, respectively. We obtain analytical forms of gauge-invariant quantities such as the absolute value of the transition dipoles $\abs{d_{a}^\kk}$, the (out-of-plane) Berry curvature $\Omega_{\pm}^\kk$ and the shift vectors $R_{ab}^\kk$ ($a=x,y$ and $b=x,y$ specify a particular Cartesian vector component)
\begin{subequations}
\label{eq:ham_4}
\begin{align}
	\abs{d_{a}^\kk}
	= & \abs{\bra{u_+^\kk}\partial_{a}\ket{u_-^\kk}}
	= \tfrac{1}{4} 
	\sqrt{ (\partial_{a} \phi_\kk)^2  \sin^2\theta_\kk + (\partial_{a} \theta_\kk)^2 } \\
	\Omega_{\pm}^\kk
	= & - 2 \Im \braket{\partial_{x} u_{\pm}^\kk}{\partial_{y} u_{\pm}^\kk}
	= \pm \tfrac{1}{2} \sin \theta_\kk [
	\partial_{x}\phi_\kk \partial_{y}\theta_\kk - \partial_{x}\theta_\kk \partial_{y}\phi_\kk ] \\
	R_{ab}^{\kk}
	= & \Delta\mathcal{A}^\kk_a - \partial_a \arg(d^{\kk}_b)
	= \partial_a \phi_\kk  \cos\theta_\kk
	- \partial_a \arg
	\Big[ \partial_b \phi_\kk  \sin\theta_\kk + i\partial_b \theta_\kk  \Big] \label{eq:ham_4c},
\end{align}
\end{subequations}
where we have used the notation $\partial_a\equiv \partial/\partial k_a$, $a=x,y$ for the partial derivates. For brevity, here we have not fully expanded Eq.~\eqref{eq:ham_4c}. 

\noindent{\bf 2. Gauge-invariant dynamics}

We implemented a gauge-invariant formulation of the semiconductor Bloch equations (GI-SBEs) \cite{GISBEs} to simulate the dynamics of the TSSs driven by intense, linearly polarized electric fields. In such a formalism, each term in the GI-SBEs is gauge invariant, which allows us to turn on/off certain properties such as dipole magnitudes and shift vectors to observe their contributions to the dynamics. We work in the dipole approximation, using a vector potential on the form $\vec{A}(t)=A(t)\uv{n} = \left[ A_{\text{MIR}}(t) +  A_{\text{THz}}(t) \right] \uv{n}$, with $\uv{n}$ the polarization unit vector. The electric field is then obtained by $\vec{F}(t) = - \dot{\vec{A}}(t) = F(t) \uv{n}$. 
In a frame moving with the vector potential $\vec{A}(t)$, or equivalently, in a basis spanned by the Houston states \cite{Yue2022JOSABtutorial, Houston1940PR}, the two-band GI-SBEs take on the form
\begin{subequations}
\label{eq:dyn_1}
\begin{align}
	\dot{f}_{\pm}^\KK(t) = & \mp 2 F(t) |d^{\KK+\aa}|\Im[P^\KK(t)] - [f_{\pm}^\KK(t) - f_{0}^\KK]/T_1 \\
	\dot{P}^\KK(t) = & - i \left[\Delta E^{\KK+\aa}  + \vec{F}(t) \cdot \vec{R}^{\KK+\aa}  \right] P^\KK(t) \notag \\
	& + i F(t) |d^{\KK+\aa}| \Delta f^\KK(t)
	- P^\KK/T_2
\end{align}
\end{subequations}
with $f_{\pm}^\KK$ the band populations (diagonal elements of the density matrix $f_{\pm}^\KK\equiv \rho_{\pm \pm}^\KK$), $P^\KK$ the band coherences (off-diagonal parts of the density matrix $P^\KK\equiv \rho_{+-}^\KK$),  $\Delta E^\kk \equiv E_+^\kk-E_-^\kk$ the band gap, $\Delta f^\KK \equiv f_+^\KK - f_-^\KK$ the population difference, $d^\kk\equiv \uv{n}\cdot \vec{d}^\kk $ the dipole along $\uv{n}$, $\vec{R}^\kk$ the shift vector ``along'' $\uv{n}$, i.e.,
\begin{align}
\label{eq:dyn_1b}
\vec{R}^\kk =
\begin{cases}
	\vec{R}^{\kk,x} \equiv (R_{xx}^\kk, R_{yx}^\kk) \qquad \text{for } \uv{n}=\uv{x} \\
	\vec{R}^{\kk,y} \equiv (R_{xy}^\kk, R_{yy}^\kk) \qquad \text{for } \uv{n}=\uv{y}.
\end{cases}
\end{align}
Phenomenological dephasing and decay times within the relaxation-time approximation, $T_2=10$ fs and $T_1=1$ ps, are included to mimic elastic and inelastic electron scattering \cite{HuberTIBallAcc}. The initial condition is chosen as $P^\KK(0) = 0$ and $f_{\pm}^\KK(0) = f_0^\KK$, where $f_0^\KK$ is the Fermi-Dirac distribution at 300 K and at a Fermi level of 0.25 eV above the Dirac point.

The total current along the Cartesian direction $a$ is
\begin{equation}
\label{eq:dyn_2}
\begin{aligned}
	j_a(t)
	= & -\sum_{\KK} \partial_aE_+^{\KK+\aa}f_+^\KK(t) - \partial_aE_-^{\KK+\aa}f_-^\KK(t) \\
	& - 2 \sum_{\KK}\Delta E^{\KK+\aa} |d_{a}^{\KK+\aa}| \Im{P^\KK e^{-i [\arg(d_{a}^{\KK+\aa}) - \arg(d^{\KK+\aa})]}}
\end{aligned}
\end{equation}
and the harmonic intensity along a specific polarization $\uv{s}$ is then calculated as $S_{\uv{s}}(\omega) \propto \omega^2 |\tilde{\vec{j}}(\omega)\cdot \uv{s}|$, where $\tilde{\vec{j}}(\omega)$ is the Fourier transform of the current $\vec{j}(t)$ weighted with a $\cos^2$ mask.
In addition, the contribution from the anomalous current depending on the Berry curvature can be calculated as
\begin{equation}
\label{eq:dyn_3}
\begin{aligned}
	\vec{j}^{\text{anom}}(t)
	= - \vec{F}(t) \times \sum_{\KK} \left[\vec{\Omega}_-^{\KK+\aa} f_-^{\KK}(t) + \vec{\Omega}_+^{\KK+\aa} f_+^\KK(t) \right].
\end{aligned}
\end{equation}

For convenience, in the main text, when the Berry curvature and shift vector are introduced conceptually, we omit their explicit $\kk$-dependence and denote them as $\vec{\Omega}$ and $\vec{R}$, provided no ambiguity arises.\newline

\section*{Data availability}
Source data for Figures 3-6 in the main manuscript are provided with this paper.

\section*{Code availability}
The codes for calculating HHG from topological surface states of Bi$_2$Se$_3$ are available upon reasonable request.

\section*{Acknowledgments}
This material is based upon work supported by the US Air Force Office of Scientific Research under Grant No.\,FA9550-2-110415 and US Department of Energy BES under Grant No.\,DE-FG02-04ER15614. A.\,S.\,L. and L.\,Y. acknowledge support from the Center for Emergent Materials, an NSF MRSEC, under Grant No.\,DMR-2011876. W.\,Z., K.\,A.\,I. and R.\,K.\,K. acknowledge support from the US Department of Energy under Grant No.\,DE-SC0016379 and the AFOSR MURI 2D MAGIC under Grant No.\,FA9550-19-1-0390. L.\,Y. acknowledges startup funds from Binghamton University.

\section*{Author contributions}
S.\,L., L.\,Y. and W.\,Z. conceived the study. S.\,L., T.\,A.\,T. and L.\,B. performed the experiment. L.\,Y. performed the simulations. K.\,A.\,I. and W.\,Z. prepared and characterized the samples. S.\,L., Y.\,T. and V.\,L. developed the laser and light source systems. Y.\,T. developed the data acquisition software packages. P.\,A., R.\,K.\,K., A.\,S.\,L. and L.\,F.\,D. supervised the study. All authors contributed to the manuscript. 

\section*{Competing interests}
The authors declare no competing interests.

\section*{Additional information}
Correspondence and request for materials should be addressed to S.L. (li.8187@osu.edu) or L.Y. (lyue2@binghamton.edu).

\newpage

\newpage

\newpage

\newpage

\newpage

\newpage

\end{document}


\title{Disentangling High Harmonic Generation from\\
	Surface and Bulk States of a Topological Insulator\\
	Supplementary Information}

\author{Sha Li}
\affiliation{Department of Physics, The Ohio State University, Columbus, Ohio 43210, USA}

\author{Wenyi Zhou}
\affiliation{Department of Physics, The Ohio State University, Columbus, Ohio 43210, USA}

\author{Kazi A. Imroz}
\affiliation{Department of Physics, The Ohio State University, Columbus, Ohio 43210, USA}

\author{Yaguo Tang}
\affiliation{Department of Physics, The Ohio State University, Columbus, Ohio 43210, USA}

\author{Tiana A. Townsend}
\affiliation{Department of Physics, The Ohio State University, Columbus, Ohio 43210, USA}

\author{Vyacheslav Leshchenko}
\affiliation{Department of Physics, The Ohio State University, Columbus, Ohio 43210, USA}

\author{Larissa Boie}
\affiliation{Department of Physics, The Ohio State University, Columbus, Ohio 43210, USA}

\author{Pierre Agostini}
\affiliation{Department of Physics, The Ohio State University, Columbus, Ohio 43210, USA}

\author{Alexandra S. Landsman}
\affiliation{Department of Physics, The Ohio State University, Columbus, Ohio 43210, USA}

\author{Roland K. Kawakami}
\affiliation{Department of Physics, The Ohio State University, Columbus, Ohio 43210, USA}

\author{Lun Yue}
\affiliation{Department of Physics, Binghamton University, Binghamton, New York 13902, USA}

\author{Louis F. DiMauro}
\affiliation{Department of Physics, The Ohio State University, Columbus, Ohio 43210, USA}


\maketitle
\tableofcontents

\newpage
\section{Thickness-dependent efficiency of surface HHG in Bi$_2$Se$_3$ thin films}
In this section, we present a detailed explanation of the thickness-dependent surface HHG in Bi$_2$Se$_3$ and the substantial enhancement observed in ultrathin films below 10 nm. 

In an ideal, free-standing Bi$_2$Se$_3$ film crystal with slab geometry, the two surfaces\textemdash top and bottom\textemdash are structurally identical. The combined inversion ($\mathcal P$-) and time-reversal ($\mathcal T$-) symmetries enforce that the spin, Berry curvature, and shift vector contributions from the two surfaces cancel each other, thereby preserving the overall $\mathcal {PT}$-symmetry of the entire slab \cite{SlabBiSe}. In a transmission-geometry HHG experiment, both surfaces are simultaneously probed, the even-order harmonics generated from the top and bottom surfaces are out of phase by $\pi$ and cancel, analogous to how randomly oriented polar molecules produce only odd-order harmonic radiation.

Our experimental observations, that an ultrathin (6 nm) film substantially enhances HHG from the surface states, while the bulk states dominate HHG in a thicker (50 nm) film, clearly contrast with the symmetry-based expectations. We attribute this thickness-dependent surface HHG to two key factors: 1. In a real crystal, the top and bottom surfaces are not perfectly identical, so their even-order optical responses do not completely cancel; 2. More importantly, the bulk and surface of a topological insulator (TI) are not fully isolated, bulk electrons can scatter into surface states, which suppresses the efficiency of surface HHG.

\textbf{Non-identical surfaces.} In a real film crystal, the top and bottom surfaces typically encounter different environments: one in contact with the substrate, the other with air, vacuum, or a capping layer. In our experiments, the bottom surface of the Bi$_2$Se$_3$ film rests on an Al$_2$O$_3$ substrate, where pre-growth treatment (see \textbf{Methods} in main text) ensures an atomically sharp interface and high surface smoothness. In contrast, the top surface is capped with CaF$_2$ without specific treatment, resulting in a more diffuse interface. These distinct environments can modify various aspects of the surface states, such as the band structure, Fermi level and degree of symmetry breaking. While it is difficult to quantify precisely how the two surfaces differ in the amplitude and phase of the even-order harmonic light they generate, their responses do not cancel completely. Additionally, light propagation effects must be considered in HHG. For Bi$_2$Se$_3$ films of a few to tens of nanometers thickness, the absorption of light is the dominant propagation effect. In transmission geometry, HHG from the top surface of a thick (50 nm) film is strongly attenuated due to absorption, whereas in an ultrathin (6 nm) film, both surfaces contribute to HHG. The sharper bottom interface is likely to exhibit a stronger symmetry breaking and to contribute more to even-order harmonic emission.
 
\textbf{Bulk-surface electron scattering.} Even after accounting for non-identical surfaces and light absorption, one might still expect weak even-order HHG in an ultrathin film if contributions from the two surfaces are similar in amplitude but interfere destructively. Yet, our experimental data show a (nearly) monotonic increase of even-order harmonic yield with decreasing film thickness (red curve in main text Figure 4), indicating that an additional mechanism affects the efficiency of surface HHG. We attribute this mechanism to bulk-surface electron scattering.

\begin{figure}[H]
	
	\centering 
	\resizebox{0.8\textwidth}{!}{\includegraphics{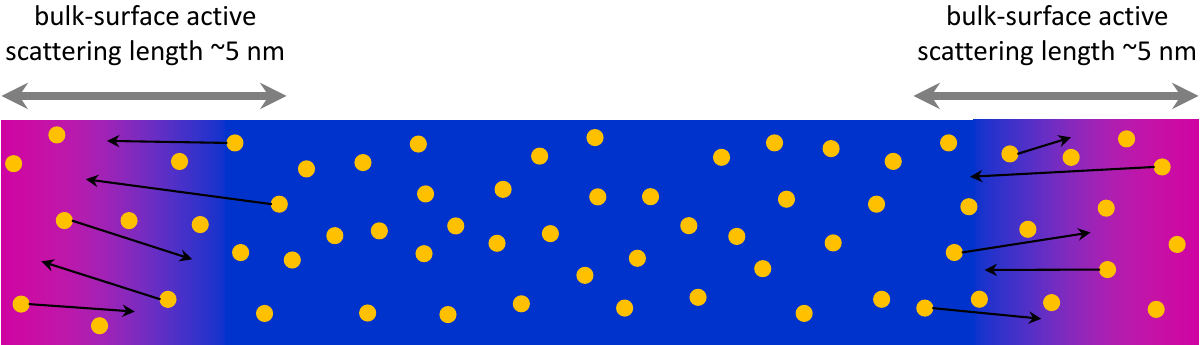}}
	\caption{\textbf{Sketch of charge transfer (via scattering) between the bulk and surface in Bi$_2$Se$_3$.} Purple: surface region, with gradient indicating decay of the surface-state wavefunction; Blue: bulk region; Orange circles: electrons. Our experimental observation suggests that the bulk-surface effective scattering length is on the order of 5 nm.}
	\label{figS1}
	
\end{figure}

Many STM and Tr-ARPES studies \cite{BSScatterSTM,BSScatterTrARPES} have shown scattering of electrons between bulk and surface states in TIs. For example, ref.~\cite{BSScatterTrARPES} reported an ultrafast increase in surface carrier density after photoexcitation, which was attributed to phonon-assisted bulk-to-surface charge transfer at an elevated lattice temperature (300 K). In general, bulk-surface electron scattering requires out-of-plane momentum transfer and thus cannot arise solely from electron dynamics driven by in-plane laser fields. Instead, it is mediated by phonons, defects, or impurities \cite{BSScatterSTM,BSScatterTrARPES}. Because bulk states are delocalized throughout the crystal while topological surface states (TSSs) are confined within only a few layers near the surface ($\sim$2 nm), only bulk electrons close enough to the surface can couple efficiently to TSSs via such scattering processes, as sketched in Figure\@ \ref{figS1}.

In trivial (gapped) semiconductors, it has been observed that photo-carrier doping incoherently generates excess hot electrons \cite{DopingHHGZnO,DopingHHG}, increasing the electron-electron scattering during HHG and thereby suppressing the harmonic generation efficiency (see the review \cite{DopingHHG} and references therein). A similar effect can occur for HHG in TSSs: photoexcited bulk electrons can scatter into surface states, leading to charge-carrier doping, while electron-electron scattering within the surface further reduces the HHG efficiency. It is important to note that in TSSs, although spin–momentum locking forbids backscattering of a single charge-carrier by time-reversal-preserving ``agents" ($e.g.$, non-magnetic impurities), it does not preclude electron-electron scattering. 

In the experiments, we observe a (nearly) monotonic increase of even-order harmonic yield as the film thickness decreases, with a steep rise below 10 nm (red curve in main text Figure 4). The sharp transition around 10 nm suggests an upper bound for the bulk-surface coupling length of about 5 nm, as sketched in Figure\@ \ref{figS1}. Therefore, below 10 nm (twice the coupling length), decreasing the film thickness enhances surface HHG because a smaller bulk volume generates less photoexcited electrons that can scatter into the surface states. Above 10 nm, the bulk-surface electron scattering becomes saturated since only bulk electrons sufficiently close to the surface couple efficiently to TSSs. In this thickness regime, the much weaker thickness dependence of surface HHG is primarily due to light absorption.  In Bi$_2$Se$_3$, the absorption lengths of the fundamental and harmonic light are respectively $>$100 nm and $\sim$20-30 nm, respectively \cite{TIabsorp}. Therefore, if the film is too thick, harmonic emission from the top surface is attenuated by absorption of harmonic light, while that from the bottom surface is suppressed due to absorption of fundamental light.

Our interpretation (of bulk-surface electron scattering) aligns with prior work. For example, in a study of HHG in Bi$_2$Te$_3$ bulk crystal in reflection geometry \cite{HuberHHGBiTe}, which in principle favors surface responses, bulk charge-carrier population was controlled by varying the photon energy of the fundamental driving laser. The absence of even-order harmonics with above-band gap driver suggested that too many bulk electrons were created and surface HHG was almost eliminated. In contrast, below-band gap driver strongly suppressed bulk excitation and enhanced surface HHG. In our study, the fundamental photon energy ($\hbar\omega_0\sim0.34$ eV) is slightly above the bulk band gap of Bi$_2$Se$_3$ ($\varepsilon_g\sim 0.3$ eV). We reduce the bulk volume and reach a similar suppression of bulk charge-carrier population. Indeed, the relative efficiency of even-order harmonic generation, quantified by the even-to-odd harmonic yield ratio, is much higher in our ultrathin (6 nm) film than those reported in previous studies. It exceeds those with above-band gap excitation of BSTS and Bi$_2$Se$_3$ bulk crystals in reflection geometry \cite{LiuHHGBSTS}, as well as that with below-band gap excitation of a 30 nm Bi$_2$Se$_3$ thick film in transmission geometry \cite{HeideHHGBiSe}. These results suggest that the reduced bulk volume in an ultrathin film plays a central role in enhancing surface HHG.

Beyond HHG, a thickness- and temperature-dependent transport study of Bi$_2$Se$_3$ films provides further support \cite{BiSeConductivity}. In that work, the contributions to the film conductivity from the surface, bulk and impurities were investigated, and it was found that at low temperature, the film conductance suddenly increased below 10 nm, attributable to the contribution from TSSs. At a thickness of 6 nm, the conduction was dominated by electrons in TSSs.

\begin{figure}[H]
	
\centering
\resizebox{0.8\textwidth}{!}{\includegraphics{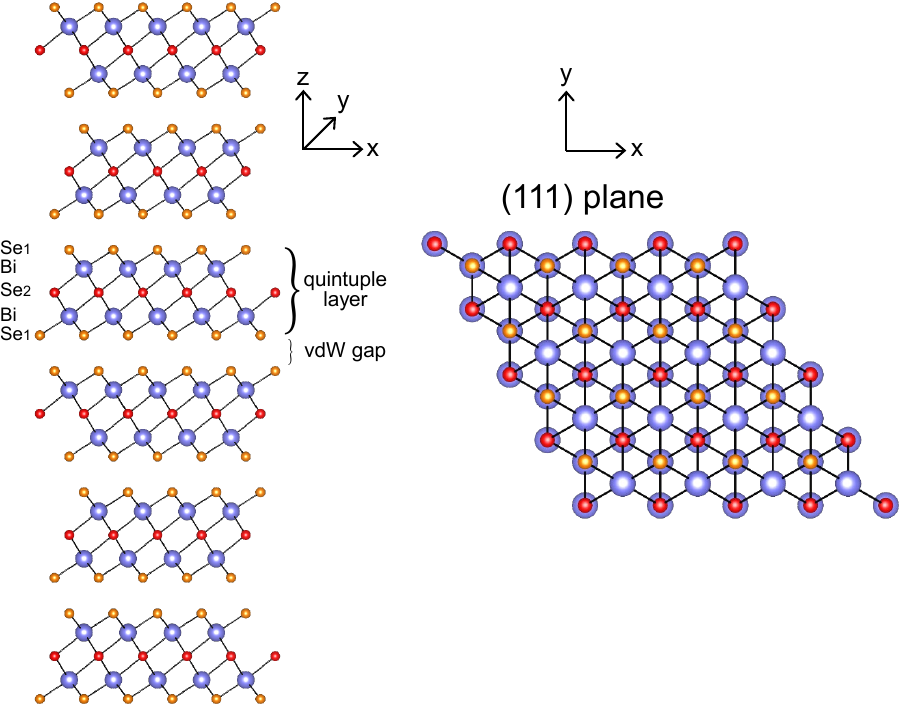}}
\caption{\textbf{Crystal structure of Bi$_2$Se$_3$.} Left: side view with the z-axis along the crystal c-axis; Right: top view of the (111) crystal plane, the x- and y-axes correspond to the reciprocal-space $\Gamma$K and $\Gamma$M directions, respectively; Blue: Bi, Yellow: Se1, Red: Se2. A quintuple layer (QL) consists of five atom layers in the sequence Se1-Bi-Se2-Bi-Se1. The QLs are held together by van der Walls forces. Inversion symmetry is preserved for both odd and even number of QLs. In odd-layer films, the inversion center is located in the Se2 atomic plane of the central QL, whereas in even-layer films, it lies within the middle van der Waals gap. The images are generated using VESTA.}
\label{figS2}
	
\end{figure}

\textbf{Excluding alternative explanations.} In the end, we rule out two possible mechanisms that may lead to thickness-dependent HHG: 1. {\it Thin-film interference} modifies the fundamental electric field inside the material that drives the HHG process, as well as the transmission of both fundamental (purple crosses in main text Figure 4) and harmonic light. However, interference would affect odd- and even-order harmonics in a similar manner, which is inconsistent with our observations. 2. {\it Layer-dependent symmetry breaking}. In a previous study of HHG in ultrathin films of transition metal dichalcogenides (TMDCs) \cite{HHGTMDstack}, the number of stacking layers was found to alter the crystal symmetry and determine whether even-order harmonic emission is allowed. Natural TMDCs typically crystallize in the 2H-phase with an ABAB... stacking order, belonging to the centrosymmetric $D_{3d}$ point group. In few-layer samples, however, the symmetry depends on the number of layers: inversion symmetry is preserved in even-layer systems ($D_{3d}$), but broken in odd-layer systems (free-standing: $D_{3h}$, on substrate: $C_{3v}$), including monolayer. In contrast, topological insulators of the Bi$_2$Se$_3$ family adopt an ABCABC... stacking sequence. The overall inversion symmetry ($D_{3d}$) is preserved regardless of the number of (quintuple) layers (Figure\@ \ref{figS2}). Consequently, in few-layer samples, symmetry breaking arises from surface or interface effects, while the number of stacking layers plays a negligible role. This is supported by our observation of a monotonic change in the even-order harmonic yield across three consecutive film thicknesses of 6, 7, and 8 nm (1 nm $\sim$ one quintuple layer (QL) = five atomic layers, the bulk lattice repeats every three QLs along the stacking direction, $i.e.$ the c-axis).

\newpage

\newpage
\section{Additional measurements: 5$^\mathrm{th}$-order harmonic generation in the 6 nm film}
\begin{figure}[H]
	
\centering
\resizebox{\textwidth}{!}{\includegraphics{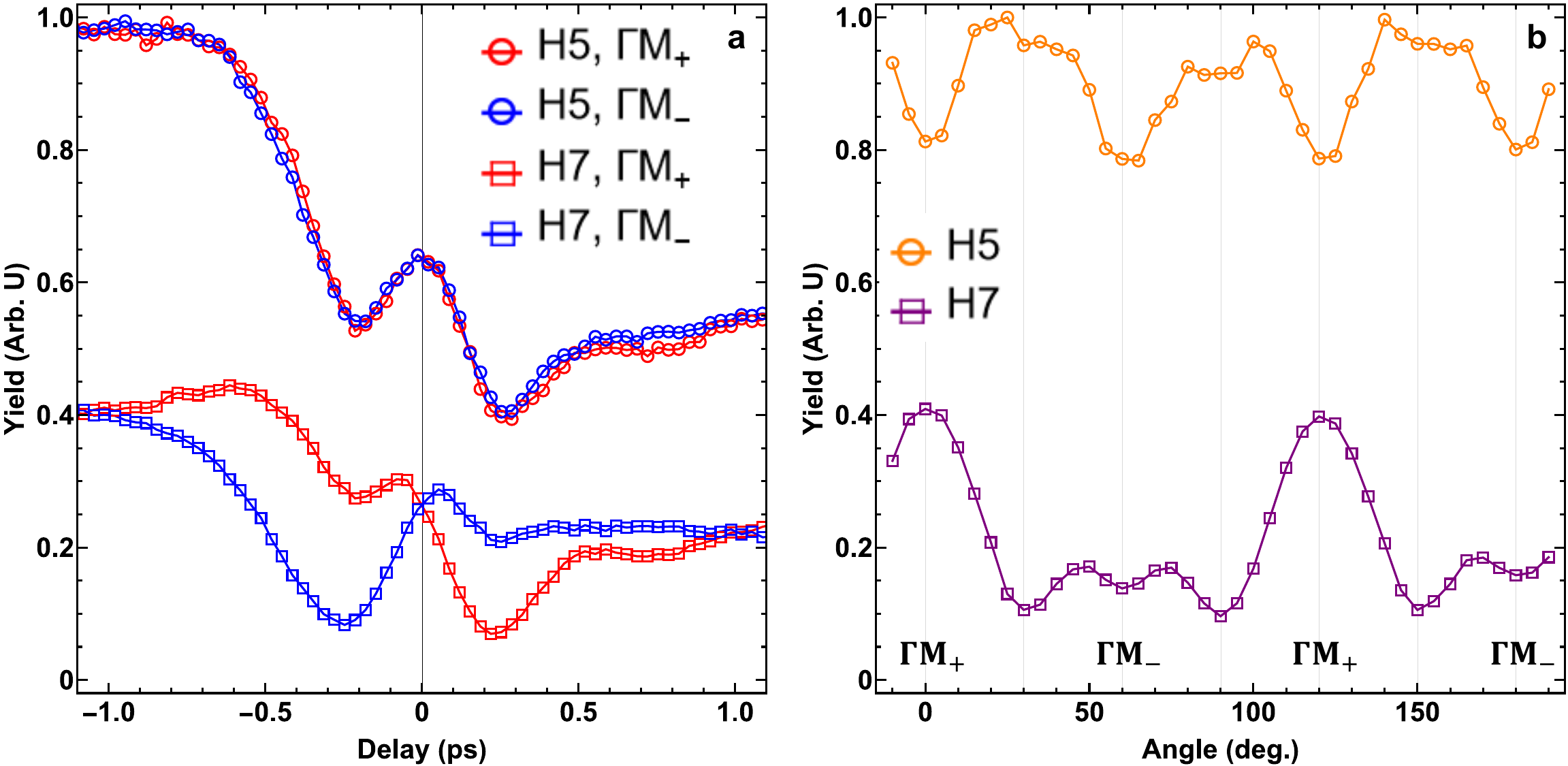}}
\caption{\textbf{Comparison of the 5$^\mathrm{th}$-order (bulk-dominated) and 7$^\mathrm{th}$-order (surface-dominated) harmonics from the 6 nm Bi$_2$Se$_3$ film.} ({\bf a}) Harmonic yield as a function of the MIR-THz time delay, showing negligible differences between the two crystal orientations ($\Gamma$M$_+$ and $\Gamma$M$_-$) for the 5$^\mathrm{th}$-order harmonic, but a large variation for the 7$^\mathrm{th}$-order harmonic. ({\bf b}) Harmonic yield as a function of the crystal orientation, measured in the presence of the THz field at a fixed MIR-THz time delay. The 5$^\mathrm{th}$- and 7$^\mathrm{th}$-order harmonics exhibit 6-fold and 3-fold symmetries, respectively. The measurements in (a) and (b) are consistent with bulk- and surface-dominated emission for the 5$^\mathrm{th}$- and 7$^\mathrm{th}$-order harmonics, respectively.}
\label{figS3}	
	
\end{figure}

In the main text, we have shown that at a fixed MIR-THz time delay, the contrast value for the 5$^\mathrm{th}$-order harmonic is extremely small (main text Figure 5c), indicating that it is bulk-dominated. This result is counterintuitive for an ultrathin (6 nm) film with minimum bulk volume. To further support our interpretation, we performed additional HHG measurements by varying the MIR-THz time delay and by rotating the crystal orientation, see Figure\@ \ref{figS3} for details. Both measurements consistently support that, even in the 6 nm film, bulk states dominate the emission of the 5$^\mathrm{th}$-order harmonic.

\newpage
\section{Roles of Berry curvature and shift vector in HHG}
In this section, we briefly discuss the roles of Berry curvature ($\mathbf \Omega$) and shift vector ($\mathbf R$) in strong-field HHG.

Recent theoretical studies have shown the importance of including the Berry connection ($\bm{\mathcal{A}}$) and transition dipole phase ($\phi$) in the semiconductor Bloch equations (SBEs) to calculate HHG in systems lacking $\mathcal P$- or $\mathcal T$-symmetry \cite{CDLinTDPHHG,BCHHG,LYHHGgauges}. While $\bm{\mathcal{A}}$ and $\phi$ are not gauge invariant, two gauge invariant vector quantities are associated with them, namely the Berry curvature ($\mathbf \Omega$) and the shift vector ($\mathbf R$). If a system preserves both $\mathcal P$- and $\mathcal T$-symmetry, $\mathbf \Omega$ and $\mathbf R$ vanish. Conversely, if at least one of the symmetries is broken, $\mathbf \Omega$ and $\mathbf R$ are generally non-zero.

$\bm{\mathcal{A}}$, $\mathbf \Omega$, $\phi$, and $\mathbf R$ are closely connected to the Berry phase $\gamma$ \cite{BerryPhase}, a fundamental concept in physics that describes the geometric phase a system acquires during an adiabatic cyclic motion, and illustrates how geometry and topology emerge in quantum mechanics. $\bm{\mathcal{A}}$ and $\mathbf \Omega$ are the local gauge potential and field associated with $\gamma$, respectively: $\gamma = \oint_c \bm{\mathcal{A}} (\kk) d\kk$ and $\mathbf{\Omega} = \nabla_\kk \times \bm{\mathcal{A}} (\kk)$. $\phi$ accounts for the phase factor of the complex transition dipole matrix (TDM) elements. Mathematically, $\bm{\mathcal A}$ and $\phi$ are the diagonal ($m=n$) elements and phase of the off-diagonal ($m \neq n$) elements of the TDM $d_{mn}^\kk=i\bra{u_m^\kk}\nabla_\kk\ket{u_n^\kk}$, respectively. Here $u_n^\kk$ is the periodic part of the Bloch wavefunction. $\mathbf{R}$ describes the shift of an electron wavepacket in real space during a transition from band $n$ to $m$, and is given by $\mathbf{R}_{mn} \equiv \bm{\mathcal A}_m (\kk)-\bm{\mathcal A}_n (\kk) - \nabla_\kk{\phi}_{mn} (\kk)$.

Non-zero $\mathbf \Omega$ and $\mathbf R$ originating from $\mathcal P$-violation are crucial for the generation of even-order harmonics. To understand this, we introduce the concept of {\it intra-cycle HHG interferometer}: The total current (or dipole) associated with the harmonic emission is the sum of currents ($\bf{j}$) from two adjacent (``$+$" and ``$-$") laser half-cycles: $\bf{j}(\omega)=\bf{j}^{+}(\omega) + \bf{j}^{-}(\omega)e^{i(\omega\Delta t + \Delta\phi)}$, and the harmonic yield $Y(\omega) \propto |\bf{j}(\omega)|^2$, here $\omega$ is the harmonic frequency. The phase difference $(\omega\Delta t + \Delta\phi)$ between $\bf{j}^+$ and $\bf{j}^-$ has two contributions: $\Delta t$ is the time separation between the two neighboring harmonic light emission events, and $\Delta\phi$ is the difference of the charge-carrier accumulated phase (semi-classical action, Berry phase, {\it etc.}). For a $\mathcal{P}$-invariant medium driven by a monochromatic laser field $\bf{j}^{+}(\omega)=-\bf{j}^{-}(\omega)$, $\Delta t = T_0/2$ and $ \Delta\phi =0$, meaning that the intra-cycle HHG interferometer is balanced to forbid even-order ($\omega=2n\omega_0$, $n \in N$) harmonic emission. Here $T_0$ and $\omega_0$ are the period and frequency of the fundamental driving field, respectively.

{\bf Berry curvature:} The Berry curvature $\mathbf \Omega$ is the local gauge field associated with the Berry phase. Analogous to a magnetic field, it can induce transverse motion of the charge-carriers. The non-zero $\mathbf \Omega$ due to $\mathcal T$-violation ({\it e.g.}, intrinsic magnetization) gives rise to the anomalous Hall effect. In the context of strong-field HHG, non-zero $\mathbf \Omega$ due to $\mathcal P$-violation induces an anomalous (perpendicular) current $\abs{\bf{j}_{\perp,\Omega}} \propto \mathbf{F} \times \mathbf{\Omega}$, leading to perpendicular harmonic emission. Furthermore, with $\mathcal T$-invariance and $\mathcal P$-violation, $\mathbf \Omega$ is odd in momentum space, {\it i.e.}, $\mathbf{\Omega} ^{\kk}=-\mathbf{\Omega}^{-\kk}$, this results in even-order only perpendicular harmonics because $\bf{j}_{\perp,\Omega}^{+}(\omega) = \bf{j}_{\perp,\Omega}^{-}(\omega)$, $\Delta t = T_0/2$ and $ \Delta\phi =0$ for the intra-cycle HHG interferometer.

{\bf Shift vector:} In a $\mathcal P$-violated medium (except for the O point group), the charge centers for different bands do not overlap in real space, resulting in a shift of the electron wavepacket during a light-induced interband transition, {\it i.e.}, the charge carrier hops from one center to another in real space. This effect is characterized by the shift vector $\mathbf R$. $\mathbf R$ is a polar vector with the unit of length, and is even in momentum space $\mathbf R^{\kk}=\mathbf R^{-\kk}$ under $\mathcal T$-invariance. It enables the generation of a net direct current (DC) upon laser excitation, namely the shift current, contributing to the bulk photovoltaic effect. For the strong-field HHG dynamics, $\mathbf R$ has several effects: 1. It effectively modifies the tunneling barrier \cite{ShiftVecTunnel}. Depending on the sign of the laser field relative to that of $\mathbf R$, the barrier is lengthened (same sign) or shortened (opposite sign), thereby suppressing or enhancing the excitation rate, which leads to $|\bf{j}^{+}(\omega)| \neq |\bf{j}^{-}(\omega)|$. 2. During charge-carrier propagation within a single band, $\mathbf R$ contributes to the accumulated phase through the term $\int \mathbf{F}(\tau) \cdot \mathbf R(\tau) d\tau$, giving rise to a non-zero phase difference between adjacent laser half-cycles. Therefore, $\mathbf R$ unbalances the intra-cycle HHG interferometer via modifying both the amplitude and phase of $\bf{j}^{\pm}(\omega)$, enabling even-order harmonic emission.

The intra-cycle HHG interferometer also explains the THz-field-induced symmetry breaking by analyzing the THz-field-induced modulations of the tunneling exciting rate, time separation ($\Delta t$) and accumulated phase difference ($\Delta \phi$) between adjacent laser half-cycles \cite{MIRTHzHHG}. Because both the (instantaneous) THz field and the shift vector are polar vectors, their coupling and the resulting impact on HHG depend on whether they are parallel (same direction) or anti-parallel (opposite directions). This allows us to differentiate between the $\Gamma \mathrm{M}_+$ and $\Gamma \mathrm{M}_-$ crystal orientations of the TSSs in Bi$_2$Se$_3$, as demonstrated in the main text.

\newpage
\section{Simulation results of HHG in TSSs of Bi$_2$Se$_3$}
In this section, we present the simulation results of HHG in TSSs of Bi$_2$Se$_3$. The simulation details are provided in the ``TSSs model and HHG simulation methods" subsection of the {\bf Methods} section in the main text. Our aim is to capture the essential features of how the shift vector and Berry curvature affect HHG, as well as how the coupling between intrinsic and THz-field-induced symmetry breaking reveals these effects. Atomic units are used unless otherwise specified.

\begin{figure}[H]
	\centering
	\includegraphics[width=0.5\textwidth, clip, trim=0 0cm 0 0cm]{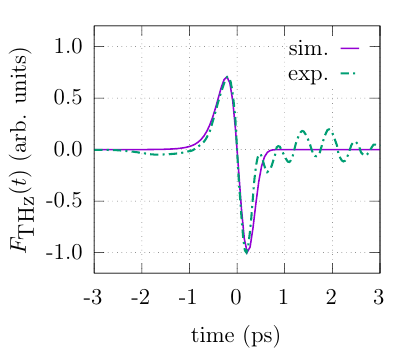}
	\caption{\textbf{THz pulse used in the simulations compared to the experimental THz waveform.}} 
\label{fig:dyn_1}
\end{figure}

We choose the MIR vector potential as $A_{\text{MIR}}(t) =A_0\cos^2[\pi t/(2\tau)]\sin(\omega_0t)$, with $t\in[-\tau, \tau]$, the carrier frequency  $\omega_0=0.0127$ corresponding to a MIR wavelength $\lambda=3.6$ $\mu$m,  the full width at half maximum (FWHM) $\tau=3475$ corresponding to a 60 fs intensity FWHM of the MIR pulse, the maximum $A_0=0.0789$ corresponding to a peak MIR electric field of 5.14 MV/cm inside the material. The THz waveform is chosen as an asymmetric single-cycle field \cite{Yang2014PRA} that matches the main single-cycle of the measured waveform, as shown in Figure \ref{fig:dyn_1}. In the experiments, the maximum achievable instantaneous THz field  inside the material is about 200 kV/cm 300 kV/cm at the peaks of its first and second main half-cycles, respectively. We propagate Eq.\,7 in the {\bf Methods} section of the main text using a 4$^\mathrm{th}$-order Runge-Kutta method with a propagation step of $0.2$ a.u., sampling the Brillouin Zone (BZ) using a $300\times 300$ $\kk$-point mesh. 

\subsection{HHG by the MIR driver alone: Roles of shift vector and Berry curvature} \label{sec:mir}

\begin{figure}[H]
	\centering
	\includegraphics[width=\textwidth, clip, trim=0 0cm 0 0cm]{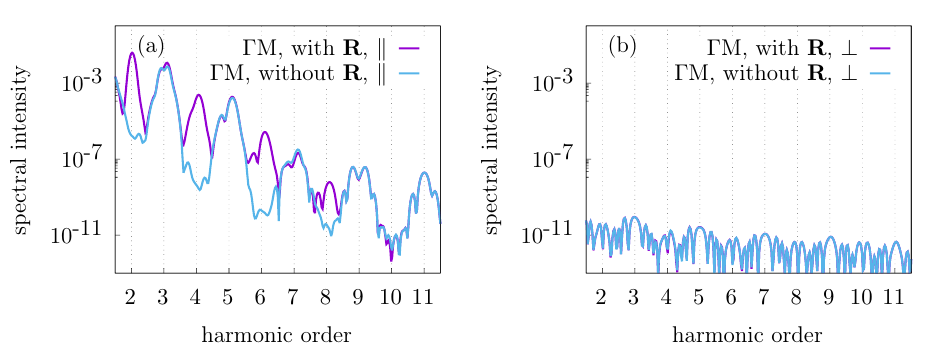}
	\caption{\textbf{Simulated HHG spectra driven by the MIR pulse polarized along $\Gamma$M}, for harmonic light polarized parallel ({\bf a}) and perpendicular ({\bf b}) to the driving laser field. Purple and blue are with and without the shift vector included in the the GI-SBEs, respectively. The orders of magnitude weaker perpendicular harmonics are due to numerical noise.} 
\label{fig:mir_1}
\end{figure}

\begin{figure}[H]
	\centering
	\includegraphics[width=\textwidth, clip, trim=0 0cm 0 0cm]{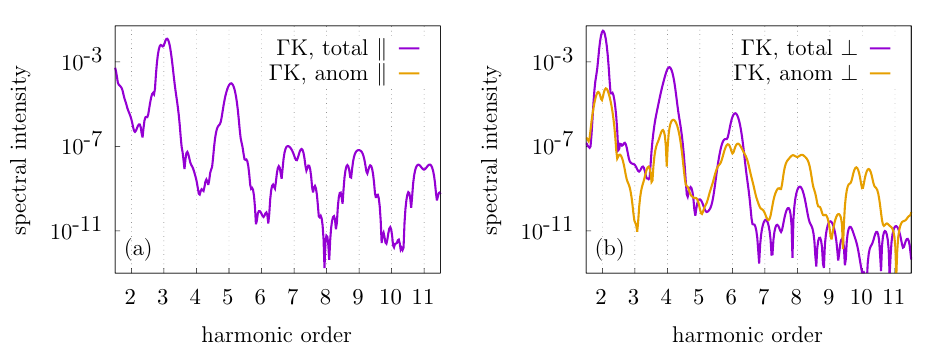}
	\caption{\textbf{Simulated HHG spectra driven by the MIR pulse polarized along $\Gamma$K}, for harmonic light polarized parallel ({\bf a}) and perpendicular ({\bf b}) to the driving laser field. Total and anomalous (Berry curvature) harmonics are plotted in purple and orange, respectively. For parallel harmonics, the anomalous contribution is 0.} 
\label{fig:mir_2}
\end{figure}

With the MIR driver alone polarized along the $\Gamma$M or $\Gamma$K direction, we calculate the HHG spectra and reproduce the symmetry-constrained harmonic polarization selection rules observed in the experiments (main text Figure 3, Figure\@ \ref{fig:mir_1}, Figure\@ \ref{fig:mir_2}). Moreover, we demonstrate how the shift vector (Figure\@ \ref{fig:mir_1}) and Berry curvature (Figure\@ \ref{fig:mir_2}) enable the generation of even-order harmonics.

In the following, unless otherwise specified, “parallel” and “perpendicular” harmonics are defined with respect to the polarization of the MIR driving field. 

Figure~\ref{fig:mir_1} shows the HHG spectra for the TSSs driven by the laser pulse polarized along $\Gamma$M. Only parallel harmonics (Figure~\ref{fig:mir_1}\textcolor{blue}{a}) are produced in this case. The perpendicular harmonic spectra shown in Figure~\ref{fig:mir_1}\textcolor{blue}{b} are due to numerical noise. Importantly, the GI-SBEs contain only gauge-invariant quantities, allowing us to directly disentangle the contribution of the shift vector by turning on/off $\vec{R}$ in our simulations. As shown in Figure~\ref{fig:mir_1}\textcolor{blue}{a}, when the shift vector is excluded from the dynamics, no even-order harmonics are produced. This clearly supports our discussion in the main text, where we attribute the emission of even-order harmonics to inversion-symmetry breaking arising from the shift vector in the case when the MIR pulse is polarized along $\Gamma$M.

For the laser field polarized along $\Gamma$K, the simulations produce odd-order-only parallel (Figure \ref{fig:mir_2}\textcolor{blue}{a}) and even-order-only perpendicular (Figure \ref{fig:mir_2}\textcolor{blue}{b}) harmonics. Moreover, the perpendicular harmonics are shown to contain an anomalous current contribution originating in the Berry curvature of the TSSs, in agreement with previous studies \cite{Sato2021PRB, HuberHHGBiTe, Yue2023PRL}. Other contributions to the perpendicular harmonic emission, including the intraband, interband, and mixture currents \cite{Yue2023PRL}, can interfere with the anomalous contribution, leading to either suppression or enhancement of the harmonic yield. In our simulations, we found that this interference is complex and depends on parameters such as the dephasing time and Fermi level. In future work, we aim to develop a more realistic TSSs model beyond the two-band tight-binding approximation to more accurately calculate the relative amplitudes and phases of these different contributions.

\subsection{THz-assisted HHG: Coupling between intrinsic and THz-field-induced symmetry breaking} \label{sec:mirthz}

One of our main objectives in this study is to demonstrate how the coupling between intrinsic and THz-field-induced symmetry breaking enables us to examine the influence of the shift vector and Berry curvature on HHG. In the following, we simulate HHG in TSSs driven by the MIR pulse in the presence of a weak single-cycle THz pulse. As a demonstration, we consider the case of a fixed MIR-THz time delay of $-210$ fs with an instantaneous THz field strength of $200$ kV/cm. This delay corresponds to the MIR pulse arriving at the peak of the first main half-cycle of the THz field.

\begin{figure}[H]
	\centering
	\includegraphics[width=\textwidth, clip, trim=0 0cm 0 0cm]{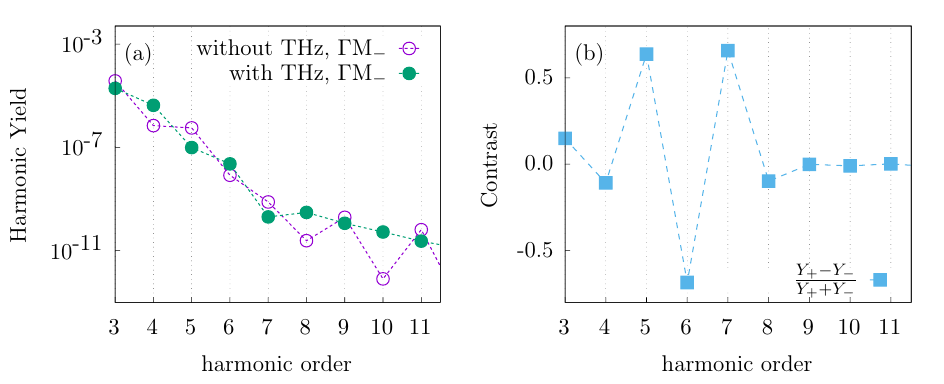}
	\caption{\textbf{THz-perturbed HHG responses for the MIR and THz fields polarized along $\Gamma$M.} ({\bf a}) Integrated harmonic yields with (green) and without (purple) the THz field oriented along $\Gamma$M$_-$. ({\bf b}) Contrast between the harmonic yields for the THz field oriented along $\Gamma$M$_+$ and $\Gamma$M$_-$.}
	\label{fig:mirthz_1}
\end{figure}

We define the THz field pointing towards or opposite to the shift vector as $\Gamma$M$_+$ or $\Gamma$M$_-$ orientations. For the MIR pulse polarized along $\Gamma$M and the instantaneous THz field along $\Gamma$M$_-$, we show in Figure \ref{fig:mirthz_1}{\color{blue} a} the harmonic yield, with and without the THz perturbation. Applying the THz field clearly enhances the emission of the even-order harmonics, while at the same time suppressing the odd-order harmonics. This is in agreement with the experimental results, see $e.g.$ the blue curves in main text Figure 5b: for the 6 nm sample, the 6$^\mathrm{th}$- (7$^\mathrm{th}$-) order harmonic is enhanced (suppressed) at delay $-210$ fs compared to at delays $<-3$ ps (MIR leads THz, but no temporal overlap).

In Figure \ref{fig:mirthz_1}{\color{blue} b}, we show the contrast between the harmonic yields for the THz field oriented along $\Gamma$M$_+$ and $\Gamma$M$_-$. Our simulations are in qualitative agreement with the experimental results shown in Figure 5c of the main text, with the odd-/even-order harmonics showing positive/negative contrast. In particular, the 6$^\mathrm{th}$- and 7$^\mathrm{th}$-order harmonics exhibit large contrasts of -0.7 and 0.6, respectively, comparable to experiment. As discussed before, the 5$^\mathrm{th}$-order harmonic is bulk-dominated, so the discrepancy in our simulation is expected since we only simulate TSSs. For harmonics above the 9$^\mathrm{th}$-order, the simulated contrast values are very small compared to experiment, reflecting the limitations of the present TSSs model and simulation method, particularly for HHG dynamics far from the $\Gamma$ point. In future work, we hope to develop ab-initio slab simulations of HHG in 3D-TI involving both the bulk and topological surface states. Nevertheless, the current simulations capture the key qualitative features of the experiment and support the interpretations presented in the main text.

\begin{figure}[H]
	
\centering
\includegraphics[width=\textwidth, clip, trim=0 0cm 0 0cm]{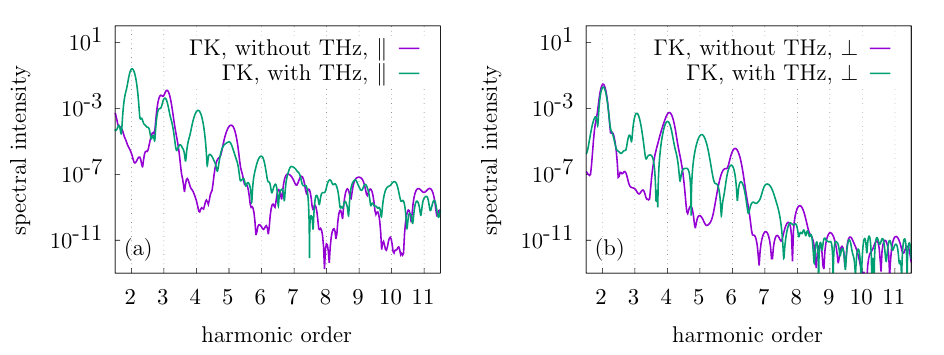}
\caption{\textbf{THz-perturbed HHG responses for the MIR and THz fields polarized along $\Gamma$K.} Parallel ({\bf a}) and Perpendicular ({\bf b}) harmonic spectra with (green) and without (purple) applying the THz field.}
\label{fig:mirthz_2}

\end{figure}

When the MIR and THz fields are polarized along the $\Gamma$K direction of the crystal, symmetry constraints enforce that perpendicular harmonics originate exclusively from the surface states. As shown in Figure \ref{fig:mirthz_2}{\color{blue} b}, with the MIR driver alone, the TSSs generate even-order-only perpendicular harmonics (purple), which contain Berry curvature contribution as discussed earlier. Introducing the THz field modulates the perpendicular harmonic spectrum (green), leading to the emergence of odd-order harmonics and a pronounced suppression of even-order harmonics, in good agreement with the experimental spectra of the 6 nm film (main text Figure 6a). The THz perturbation also modifies the parallel harmonics (Figure \ref{fig:mirthz_2}{\color{blue} a}); however, since the bulk and surface states exhibit similar responses, their contributions to the parallel harmonics cannot be readily separated. These results highlight the effectiveness of using the THz field to selectively probe surface-state HHG.

